\documentclass[preprint,
 amsmath,amssymb,
 aps,
]{revtex4-2}

\usepackage{graphicx}
\usepackage{dcolumn}
\usepackage{bm}
\usepackage{hyperref}
\usepackage{xcolor}

\usepackage{soul}

\usepackage{subfigure}

\begin{document}

\title{Loss of Axial Symmetry in Hypersonic Flows over Conical Shapes}

\author{Irmak T. ~Karpuzcu}
 \email{itk3@illinois.edu}
 \author{Deborah A. Levin}
 \affiliation{University of Illinois, Urbana-Champaign, IL 61801, USA}
\date{\today}

\begin{abstract}
The assumption of axial symmetry for hypersonic flows over conically shaped geometries is ubiquitous in both experiments and numerical simulations.  Yet depending on the free stream conditions, many of these flows are unsteady and their transition from laminar to turbulent is a three-dimensional phenomena.  Combining triple deck theory/linear stability analysis with the kinetic direct simulation Monte Carlo method we analyze  the azimuthal eigenmodes of flows over single and  double-cone configurations.   For  Mach 16 flows we find that the strongest amplification rate occurs for the non-axisymmetric  azimuthal wavenumber of $n=1$.  This occurs in regions quite close to the tip of the cone  due to the proximity of the conical shock to the viscous shear layer where non-axisymmetric modes are amplified through linear mechanisms.  Comparison of  triple deck linear stability predictions shows that in addition to the azimuthal wavenumber, both the temporal content and  amplification rate of these non-axisymmetric disturbances agree  well with the time accurate DSMC flowfield.  In addition to the loss of axial symmetry observed at the conical shock, the effect of axial symmetry assumptions on  the more complex shock-shock and shock-boundary layer interactions of a flow over a double cone are considered.   The results for the separation region show that axisymmetric and three dimensional simulations differ in almost all of the main  flow structures. Three dimensional flowfields result in a smaller separation bubble with weaker shocks and three dimensional effects  were manifest in the variation in surface parameters in the azimuthal direction as well.   Interestingly, the DSMC simulations show that  the loss of axial symmetry in the separation region, begins near the cone tip.
\end{abstract}

\maketitle



\pagebreak
\section{Introduction\label{sec:intro}}

Hypersonic flows over conical geometries are of interest as they represent simplified versions of many hypersonic vehicles.  As such, it is well recognized that many of the flows over such shapes are unsteady, and, depending on the specific free stream conditions will transition from laminar to turbulent~\citep{MackReport,Hussainibook,Schmidlinear}. The specific moment the flow transition occurs depends on the evolution which  is strictly three-dimensional in nature regardless of whether the flow geometry is axisymmetric.  However, the vast majority of literature for modeling and simulation as well as experiments does not consider the flows as truly three-dimensional. This paper explores whether three dimensional effects are important in characterizing unsteady shock-dominated hypersonic flows for conical geometries.

In terms of experiments, many researchers studied transition to turbulence on slender cones ~\citep{MADDALON1968Experimental,SOFTLEYExperimental1969,demetriades1974hypersonic,stetson1980hypersonic}. Of these, Stetson~\cite{stetson1980hypersonic} considered several configurations of nose tip bluntness, angle of attack, Reynolds numbers and Mach numbers as well as the biconic configurations. These experiments form a database of slender cone transition studies and show that transition is delayed for blunt cone cases compared to  sharp cones. Several researchers followed in their footsteps  and observed that the inviscid Mack mode is responsible for transition although a review by Schneider~\cite{SchneiderReview2004} of transition experiments for slender cones states that the main source of uncertainty in these experiments is the free stream noise levels. However, despite the large  number of experimental investigations for any conical shape, the flow is still assumed to be axisymmetric, and therefore the role of three-dimensionality in inducing transition is lost.


Biconic configurations create more complex flowfields, providing excellent test cases for numerical models that are required to capture shock - shock interactions regions such as  Edney type-IV interactions~\citep{Edney}. An example is the experimental campaign of Holden \textit{et al.}~\cite{holdenDCExp} for the 25-55$^\circ$ half angle double cone geometry. 
Experimental measurements of surface pressure and heat transfer showed that steady flow was established in the Cubric LENS shock tube for different 
free stream Mach and Reynolds numbers. In contrast, similar experiments for the biconic geometry configuration resulted in the unsteadiness of the triple point formed by shock interactions and non-linear oscillations of the flow~\citep{jagadeesh2003visualization}. There are also additional recent experimental studies that reassess
  the initial experiments by Holden \textit{et al.}~\cite{holdenDCExp}, in terms of free stream conditions, unsteadiness and three dimensionality in the flow~\citep{RayEstimation2020,RayDCassesment2023,McGilvrayDCScitech2023}. It was found in these works that the characterization of free stream conditions is critical to the established flow over the geometry as well as indications of three dimensionality in the separation region. Nevertheless, no conclusive evidence was gathered regarding the three dimensionality of the flow.


Similar to experiments, when it comes to theoretical studies informed by modeling and simulation, transition to turbulence is the main focus for slender cone geometries. In the well-known report, Mack~\cite{MackReport} finds from linear stability theory that the second mode (the inviscid mode) becomes the dominant mode compared to the first mode (viscous mode) to cause transition and the critical Reynolds number for the second mode becomes unstable as the Mach number increases. Many studies after this report investigated the effect of second mode instability on transition to turbulence for sharp and blunted cones using linear stability analysis~\citep{malik1990stability,kufner1995entropy,schneider2001hypersonic,LyttleComparison2005}.  Comparison of these linear stability results with the available experimental data showed that the linear stability characteristics of these flows depend on several parameters such as the wall temperature and heat transfer, and the fidelity of non-equilibrium modeling. The dominance of the second Mack mode was also observed in recent numerical studies as well~\citep{Jewell2017,Paredes2019NonModal,MarineauEtAl2019secondmode}. Global linear stability analysis~\citep{BhoraniyaVinod2017} and the cross flow instabilities with sharp cones at non-zero angle of attack~\citep{Feietal2016,Chenetal2023} were also considered. More recently, three dimensional receptivity of the boundary layer to the free stream disturbances was investigated for hypersonic flow over sharp and blunt cones~\citep{CookNicholsReceptivity2024}. It was found that the receptivity of the flow is highly three dimensional and amplification of the disturbances are localized to specific azimuthal angles. However, similar to the experimental work, the effect of the conical shock and the flow three dimensionality prior to transition was not heavily considered in these theoretical and numerical studies.


With respect to double cones, Harvey \textit{et al.}~\cite{harveyDCvalid} were among the first to simulate such flowfields using both molecular kinetic direct simulation Monte Carlo (DSMC)~\citep{Bird} and the Navier-Stokes equations. Test cases at M$_\infty$=9.5 and M$_\infty$=11.4  for several Reynolds numbers
 showed that DSMC surface parameters match measurements extremely well upstream of the separation region, however, predicted a smaller separation compared to the experiments. Even though experiments showed that the flow was steady, several numerical studies predicted unsteady behavior as pointed out in the review by Knight \textit{et al.}~\cite{KNIGHT20128}. Unsteadiness of the flowfield was also investigated by Tumuklu \textit{et al.}~\cite{tumukluPoF,tumukluPoF2} for the 25-55$^\circ$ configuration using the axial symmetry assumption at M$_\infty$=16 and three different free stream Reynolds numbers of $O(10^{5} m^{-1})$ using DSMC. They showed that both low-frequency  and Kelvin-Helmholtz type unsteadiness are present in the flow due to  shock-boundary layer interactions. Hao \textit{et al.}~\cite{hao_fan_cao_wen_2022} subsequently showed that the flow is three dimensional in the separation region using DNS and LSA for the same configuration by assuming an a-priori known azimuthal wavenumber in their 3D simulations. Although there are several computational studies for the biconic configuration, the three dimensional nature of the flow prior to the interaction region was not thoroughly investigated~\citep{nompelis2010numerical,Hao_Wen_2020,HornungDCAxi}.


Despite the general lack of focus on three-dimensionality effects prior to transition and the effect of the attached conical shock promoting instability, the following investigators have studied some aspects of three-dimensional instabilities. The breaking of axial symmetry in a conical shock was observed experimentally for converging shock cases at supersonic free stream conditions in Takayama \textit{et al.}~\cite{Takayama1987} where it was shown that as the shock converges to the singular point (that is representative of the tip of the cone), axial symmetry is largely lost. In addition, any defect caused by the testing facility does not decay and stays within the shock layer. Chou \& Schneider~\cite{ChouSchneider2018} showed that off-set disturbances entering through the leading edge will cause the flow to become three dimensional. The stability of the conical shock was theorized and it was shown that infinitesimal disturbances from the free stream can cause asymmetry to the shock itself~\citep{Jun2011,Dening2016,Chen2021}. In numerical studies for converging shock-waves~\citep{Murakami2015,Chefranov2021}, it was also shown that linear perturbations in shocks can grow within certain flow parameters involving cone half angles, Mach number and ratio of specific heats. Using triple deck formulations, Duck \& Hall~\cite{DuckHall21990} showed that non-axisymmetric modes are the most important ones for a supersonic flow over an axisymmetric geometry. The effect of the attached shock on the stability of flows past sharp wedges and cones was investigated by Cowley \& Hall~\cite{CowleyHall1990} and Seddougui \& Bassom~\cite{SeddouguiBassom1997} respectively, where both analyses were performed for hypersonic flows, using linearized triple deck formulations. In both of these works, it was shown that when the shock is included in the calculations for the decks, three dimensional modes are amplified more than their 2D or axisymmetric counterparts. It was also shown in Seddougui \& Bassom~\cite{SeddouguiBassom1997} that as the shock approaches
the surface of the cone, the amplification rates of the non-axisymmetric disturbances grow rapidly. The recent work of Karpuzcu \textit{et al.}~\citep{karpuzcu2024linear} showcased the contribution of the leading edge shock for a laminar separation bubble created by a compression ramp to become three dimensional through amplification of three dimensional perturbations using global linear stability analysis and DSMC. In our preliminary work concerning the stability of hypersonic flow over conical geometries~\citep{KarpuzcuSciTech2023DC,KarpuzcuRGDDC2024}, we showed that there are implications that axial symmetry is lost within the high gradient shock and shear layers of the flow over the double cone configuration prior to the interaction region. These studies suggest that given that even simpler shock configurations shown to have unsteady frequencies~\citep{sawant_POF,sawant_TCFD}, a detailed investigation of the flowfield that involves the three dimensional conical shock is necessary.

As these questions regarding the three dimensionality of flows over conical geometries persist, we combine the use of the particle-kinetic direct simulation Monte Carlo (DSMC)~\citep{Bird} with the linear stability analysis of triple deck theory (TDT) ~\citep{SeddouguiBassom1997} for the first time to investigate the  origins of three dimensional disturbances in canonical flows over conical bodies. Since the flow is hypersonic and there are several high gradient flow regions such as shocks and shear layers where we will later show that  continuum breakdown occurs, thus suggesting the use of a kinetic method to provide the highest fidelity of the inner shock structure~\citep{Bird_1967,Bird_1968,Bird1970,alsmeyer1974messung,Alsmeyer_1976}. DSMC was shown to resolve multi-scale flow phenomenon exceptionally well by a number of researchers, and it is an inherently time accurate method that provides insights to the unsteady nature of the flows~\citep{GallisDSMCvsNS,tumukluPoF,tumukluPoF2,sawant_etal_2022,GallisTurbolenza,KarpuzcuSideJet2023}. The organization of the paper is as follows; Section~\ref{sec:NumericalGeometry} summarizes the free stream conditions, geometrical configurations, and numerical parameters used in this work. In Section~\ref{sec:singleConeStability} we characterize the non-axisymmetric instabilities within the shock and shear layers over a simple sharp cone using a fully three-dimensional DSMC and linear stability triple deck analysis. Section~\ref{sec:DoubleConeStability} presents full three dimensional simulations over the double cone configuration where in this case the additional complication of the unsteady three-dimensional recirculation bubble and its role in contributing to the breaking of  axial symmetry on the second cone must be considered. Finally, a summary of the work and conclusions are given in Section~\ref{sec:conc}.

\section{Numerical approach and selection of geometry}\label{sec:NumericalGeometry}

In this study, the linear stability of a hypersonic flow at M$_\infty=16$ over two conical geometries is considered. The first geometric  configuration is a single sharp cone with a 25$^\circ$ half angle.  This geometry corresponds to the upstream portion of a double cone that has a 25-55$^\circ$ half angle double cone with a base diameter of  $0.26~m$, the second configuration. Both are modeled as fully three dimensional  using direct simulation Monte Carlo (DSMC)~\citep{Bird}.  Table~\ref{tab:FreeStreamDSMC} provides the free stream flow parameters for a molecular nitrogen working fluid at a Re$_\infty=1\times10^{5}~m^{-1}$, the smallest Reynolds number analyzed in the previous  axisymmetric work of Tumuklu \textit{et al.}~\cite{tumukluPoF,tumukluPoF2}.   
In the present work, the massively parallel SUGAR-3D DSMC solver~\citep{Sawant2018} which employs the majorant frequency scheme~\citep{majorant} for modeling collisions is used.  To accurately model the viscosity-temperature dependence, the Variable Hard Sphere (VHS)~\citep{Bird} model  was used with the species dependent parameters also given in Table~\ref{tab:FreeStreamDSMC}.  The internal modes of  molecular nitrogen were modeled but no chemical reactions were included in the simulations as the total enthalpy of the flow was deemed as too low for reactions to occur.  

In DSMC, two different cell types are used in the simulation.  The most important one, the collision cell size, must be on the order of the local mean free path to ensure a true solution of the Boltzmann equation of transport.  Grid adaptation was performed to ensure that collision cells meet that criteria and  we confirmed that there were at least 20 computational particles per collision cell to ensure valid statistics for each time step.  The second grid, known as sampling grid, is used to compute flow macroparameters in a time accurate manner.  
  Table \ref{tab:NumericalDSMC} summarizes the  DSMC computational parameters used  in this paper.   Note that for 
 results shown in the following sections macroparameters were sampled after the flow established over the geometry.  

In this work, the tips for both single and double cone configurations were considered to be sharp. The gas particle surface collisions are handled purely by the mathematical formulation of intersection of a ray with a Lipschitz cone to determine whether or not a particle is hitting the surface of the cone, using the velocity vector of the gas particles. To perform gas-surface collisions and store the surface parameters a geometry that is consisting of planar geometrical elements was used. The cone tips are single tetrahedron elements with edge lengths of $0.1$ and $0.01~mm$ for double and single cones respectively where the tip of this tetrahedron element is the tip of the cone. The sharp tips of the single and double cone configurations have an order of magnitude difference in the tip tetrahedron element sizes and they both still show the same instability structure as will be demonstrated in Section \ref{sec:singleConeStability}. Thus the effect of the sharpness of the geometry is modeled with sufficient resolution. Furthermore, although the orientation of the faces of these tetrahedron elements are also different in the double and single cone configurations, the effect of describing the curved surface with planar elements seem to have no serious consequence.

\begin{table}
\caption{Free stream conditions for the 3D simulations of single and double cone configurations}
\label{tab:FreeStreamDSMC}
\begin{center}
\scalebox{0.8}{
\begin{tabular}{lll}
\hline
\hline
Parameter           & Unit       & Value \\
\hline
Mach Number         & [ -- ]     & 16.0  \\
Temperature         & [ K ]      & 42.6  \\
Free stream velocity, $U_\infty$ & m/s & 2073.0 \\
Number Density      & [ m$^{-3}$ ] & $3.75\times10^{21}$  \\
Unit Reynolds Number & [ m$^{-1}$ ] & $1\times10^{5}$  \\
Mean free path & [ m ] & $1.89\times10^{-3}$  \\
Molecular Diameter & [ \AA ] &       4.17\\
Viscosity index & [ -- ]           & 0.745\\
\hline
\hline
\end{tabular}}
\end{center}
\end{table}

\begin{table}
\caption{DSMC computational parameters for the 3D simulations of single and double cone configurations}
\label{tab:NumericalDSMC}
\begin{center}
\scalebox{0.8}{
\begin{tabular}{cccccccccc}
\hline
\hline
Parameter           & Unit       & Single Cone & Double Cone\\
\hline
Sampling Cell Volume & [ $mm^3$ ] & $ 0.42$& 1.00 \\
Number of Sampling Cells & & $8 \times 10^6$ & $32 \times 10^6$ \\
Time Step & [ s ]& $ 2.0\times10^{-8}$ & $ 2.0\times10^{-8}$ \\
Domain Size (x,y,z) & [ m,m,m ] & 0.15,0.15,0.15 & 0.2,0.4,0.4 \\
Total number of simulation particles & [ \# ]& $ 10\times10^{9}$& $35\times10^{9}$ \\
\hline
\hline
\end{tabular}}
\end{center}
\end{table}

\begin{figure}
\center
{\includegraphics[trim=2cm 2cm 2cm 1cm,clip,width=0.60\linewidth]{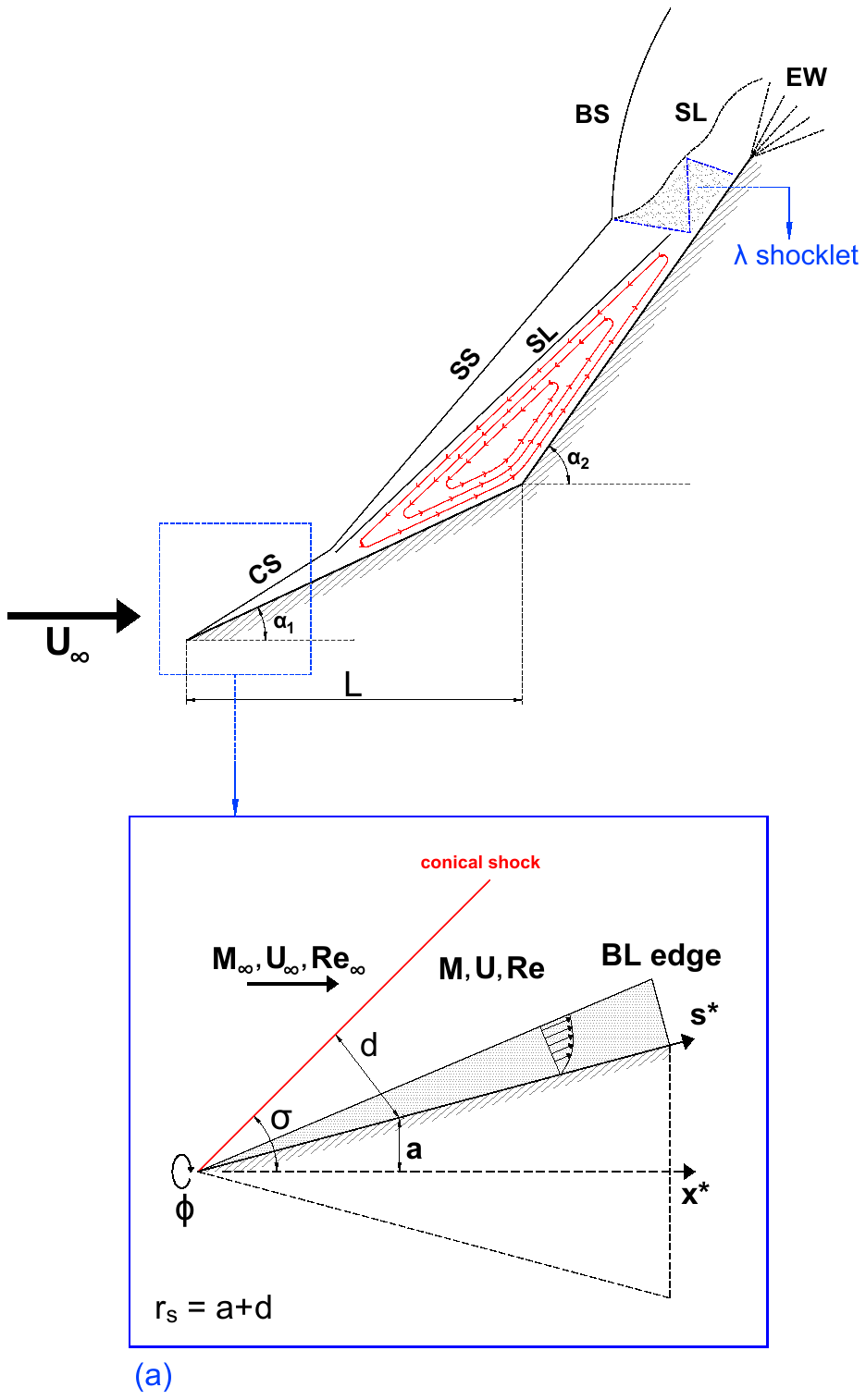}}
\caption{Overall flowfield characteristics for the double cone flows. CS: conical shock, SS: separation shock, SL: shear layer, BS: bow shock, EW: expansion waves. The detail in (a) for the region close to the tip shows the parameters that will be used in the linear stability analysis in Section.~\ref{sec:singleConeStability}.  The axial length of the first cone, $L=0.1~m$, and the tip of the first cone is the origin point in subsequent plots.  }
 \label{fig:SchematicDC}
\end{figure}

 Fig.~\ref{fig:SchematicDC}(a) shows the definitions for the conical shock geometry relative to a conically shaped body.   As is well known, a sharp cone creates an attached conical shock with a shock angle $\sigma$ from the axis of symmetry with a boundary layer starts developing from the tip.  For this single cone configuration, a higher spatial resolution was achieved compared to the double cone case with smaller sampling cell volumes and a higher number of simulation particles per sampling cell, as shown in Table~\ref{tab:NumericalDSMC}. This   higher spatial resolution for the single cone enables us to resolve the detailed structures within the shock and shear layers, especially close to the tip,  to understand their importance in causing the loss of axial symmetry. 

\section{Origin of loss of axial symmetry in a conical shock for flow over a sharp cone}\label{sec:singleConeStability}

First we consider three-dimensional simulations over a single cone to understand the origin of axial symmetry breakdown inside the leading edge shock, close to the cone tip.  
To characterize the unstable behavior of the flow in the region close to the sharp cone tip, we follow the work of Seddougui \& Bassom~\cite{SeddouguiBassom1997}  (hereafter referred to as S\&B) who used triple deck theory (TDT) with linear stability theory to study viscous instabilities for an attached conical shock. S\&B derived expressions for the axisymmetric and non-axisymmetric disturbances as,
\begin{eqnarray}
\frac{{Ai^{'}({\xi _0})}}{{\int\limits_{{\xi _0}}^\infty  {Ai(\xi )d\xi } }} & = & - {(i\alpha )^{4/3}}\frac{{{I_0}(i\alpha {r_s}){K_0}(i\alpha a) - {I_0}(i\alpha a){K_0}(i\alpha {r_s})}}{{{I_1}(i\alpha {r_s}){K_0}(i\alpha a) - {I_1}(i\alpha a){K_0}(i\alpha {r_s})}}  \label{eqn:TDTAxiCase}  \\
\frac{{Ai^{'}({\xi _0})}}{{\int\limits_{{\xi _0}}^\infty  {Ai(\xi )d\xi } }} & = & - {(i\alpha )^{4/3}}\frac{{{n^2}}}{{{\alpha ^2}{a^2}}}\frac{{{I_n}(i\alpha {r_s}){K_n}(i\alpha a) - {I_n}(i\alpha a){K_n}(i\alpha {r_s})}}{{{I_n}(i\alpha {r_s}){K_n}^{'}(i\alpha a) - {I_n}^{'}(i\alpha a){K_n}(i\alpha {r_s})}} \label{eqn:TDTNonAxiCase}
\end{eqnarray}
where $Ai$ and $I_n$ are the Airy function and n-th order Modified Bessel function of the first kind, respectively (~\cite{AbramowitzStegun}),  $r_s$ and $a$ indicate the shock location and cone surface (see Fig.~\ref{fig:SchematicDC}),  and
\begin{eqnarray}
\xi  & = & {\xi _0} + {(i\alpha )^{1/3}}Y  \\
{\xi _0} & = & - {i^{1/3}}\Omega {\alpha ^{ - 2/3}}.
\end{eqnarray}

A number of the scaling relations for the flow variables are given in S\&B based on the scaling laws of the triple deck theory~\citep{neiland1969theory,stewartson1970laminar,rizzetta1978triple},
\begin{eqnarray}
\overline a  &= & {{\mathop{\rm Re}\nolimits} ^{ - 3/8}}{\mu _w}^{3/8}{\lambda ^{ - 5/4}}{T_w}^{9/8}{M^{ - 1/4}}a \label{eqn:ScaleCone} \\
s &= &1 + {{\mathop{\rm Re}\nolimits} ^{ - 3/8}}{\mu _w}^{3/8}{\lambda ^{ - 5/4}}{T_w}^{9/8}{M^{3/4}}S \label{eqn:ScaleX} \\
t & = & {{\mathop{\rm Re}\nolimits} ^{ - 1/4}}{\mu _w}^{1/4}{\lambda ^{ - 3/2}}{T_w}^{3/4}{M^{1/2}}\tau  \label{eqn:ScaleT}
\end{eqnarray}
where $\overline a$ is the scaled local radius of the cone,  $T_w$ is the wall temperature, $\mu_w$  is the dynamic viscosity at the wall temperature, $\lambda$ is the boundary layer constant from Blasius' solution where a value of 0.332 was used, and $M$ is the Mach number. All the flow variables used in these above formulations are post-shock flow variables except for   the Reynolds number which is calculated as a local quantity, 
\begin{equation}
Re={\rho U s^*\over \mu}
\end{equation}   
For each $s^*$ value, the post shock quantities are obtained at the edge of the boundary layer.  Note that the variable $s$ represents the locally normalized distance along the surface of the cone, $S$ is the scaled distance along the surface of the cone and $x^{*}$ is the dimensional axial length in streamwise direction. Also we use the convention that dimensional variables are represented by a superscript $^{*}$.

As discussed in S\&B, this analysis is valid for locations where the boundary layer thickness is small compared to the cone radius. Due to the half angle of the cone and the high Mach number considered in this work, our case satisfies this condition starting from the immediate downstream location of the cone tip. This is because the distance of the shock to the surface is always about 10\% of the local cone radius and the boundary layer thickness does not exceed the shock.
 Furthermore, they show that this corresponds to the conditions where,
\begin{equation}
\label{eqn:ValidIntervals}
M \sim {\sigma ^{13/14}}{{\mathop{\rm Re}\nolimits} ^{3/14}},{{\mathop{\rm Re}\nolimits} ^{ - 1/9}} \le \sigma  \le {{\mathop{\rm Re}\nolimits} ^{ - 1/107}}.
\end{equation}
With $Re=2000$ at a location of $s^*=33~mm$ from the tip of the cone and the angle between the conical shock and the center axis, $\sigma=28^\circ$, our single cone configuration is within those limits. It should also be noted that $\overline a$, $s$ and $t$ are already non-dimensionalized with the appropriate combinations of the post shock speed and the distance from the tip of the cone along the cone surface for the location under consideration.

Equations~\ref{eqn:TDTAxiCase} and \ref{eqn:TDTNonAxiCase}, subject to the conditions stated above, are solved for either $\alpha$ or $\Omega$ corresponding to the spatial or temporal linear stability problem to obtain the stability characteristics of the flow assuming that the perturbations have the following form,
\begin{equation}
\label{eqn:perturbationansatz}
E = \exp [i(\alpha S + n\phi  - \Omega \tau )]
\end{equation}
where $\alpha$ is the spatial growth rate along the axial direction, $n$ is the azimuthal wavenumber, and $\Omega $ is the frequency of the disturbance. As Mack showed in  linear stability analysis \citep{MackReport}, the azimuthal modes always have integer values for axisymmetric cases, so in the following analysis $n$ always has an integer value. To determine the spatial stability modes, $\Omega$ is assumed to be real and  Eqs.~\ref{eqn:TDTAxiCase} and \ref{eqn:TDTNonAxiCase}, 
are solved for complex $\alpha$ values whereas the temporal stability modes are obtained for $\alpha $ real and  complex $\Omega $.  Note that negative values of the imaginary part of  $\alpha$ or the positive values of the imaginary part of the $\Omega$ 
indicate an unstable or growing mode.     Equations~\ref{eqn:TDTAxiCase} and \ref{eqn:TDTNonAxiCase} were solved using a Newtonian iteration solver and found to compare well with Figure~11 of S\&B~\cite{SeddouguiBassom1997} for the case of n = 1, $a$=3.0, $r_s$ = 4.0, and $\alpha$ for temporal stability and Figure~14 of S\&B for the case of n=1, $a$=0,75, $r_s$=1.0 and $\Omega$ real  for spatial stability.

\begin{figure}[htb!]
\begin{center}
\subfigure[]{\label{fig:omegaalfaix5mm}\includegraphics[trim=2cm 1.3cm 3cm 2.5cm,clip,width=0.48\linewidth]{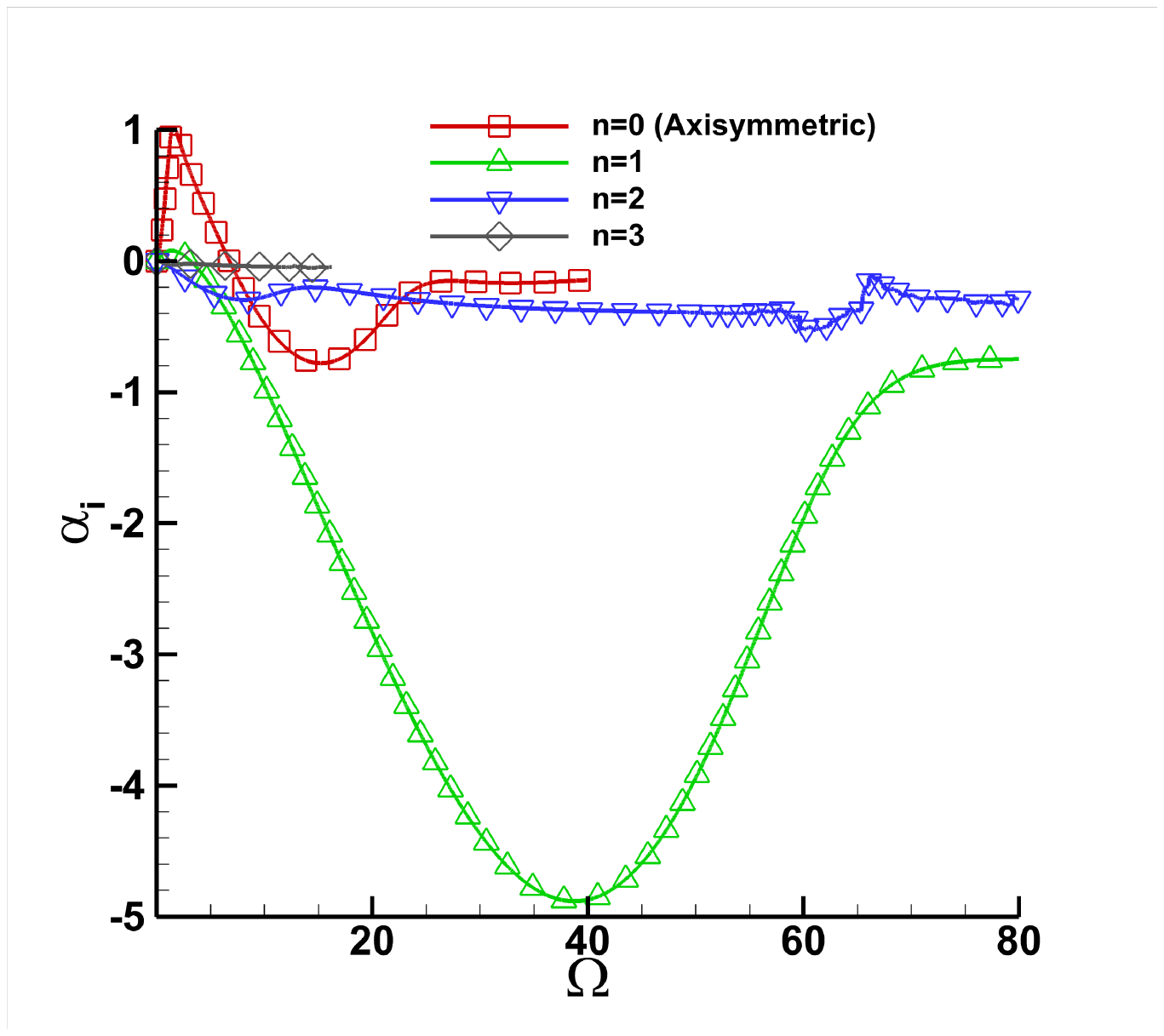}}
\subfigure[]{\label{fig:omegaalfaix30mm}\includegraphics[trim=2cm 1.3cm 3cm 2.5cm,clip,width=0.48\linewidth]{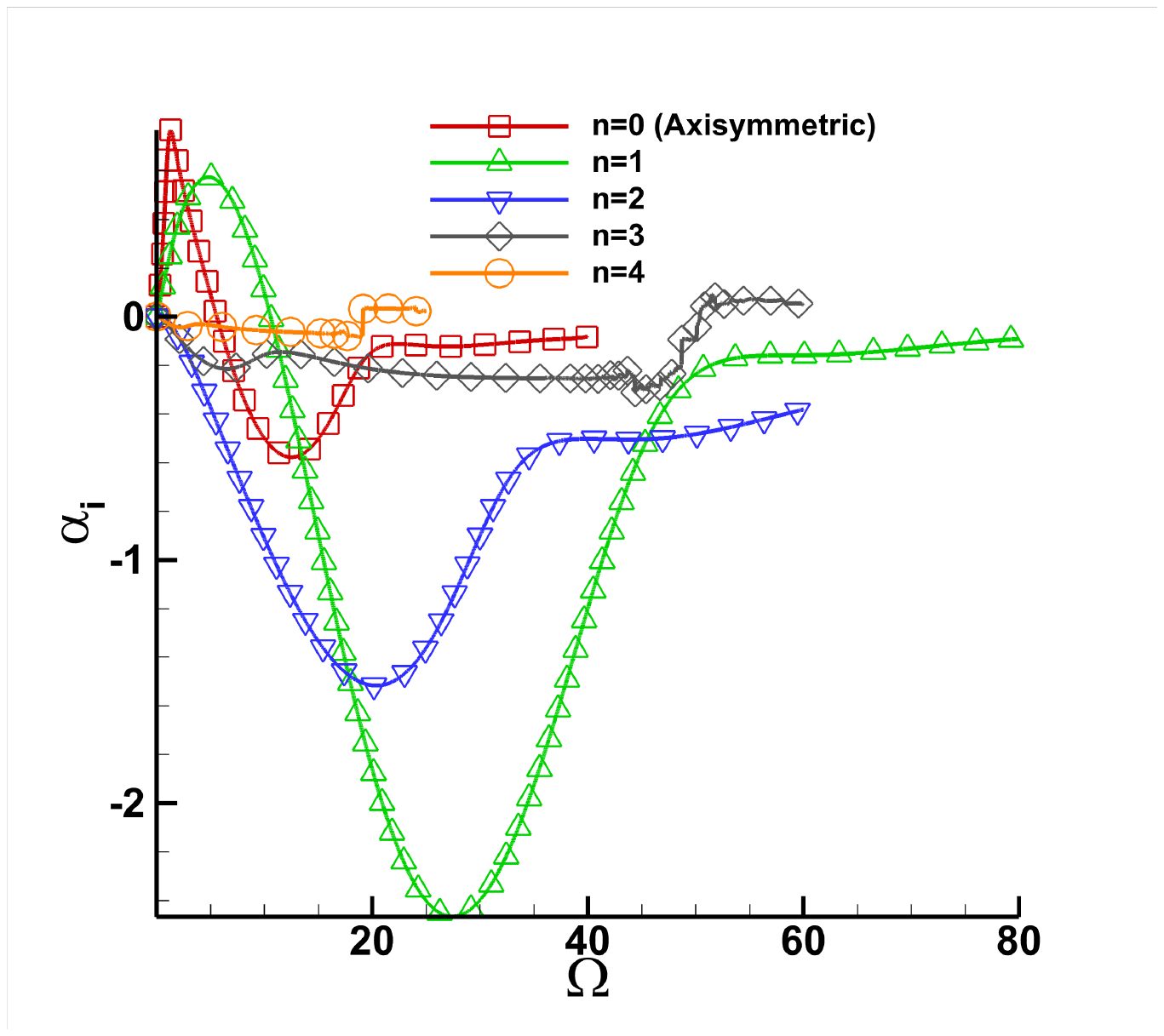}}
\end{center}
\caption{Most amplified modes for each azimuthal wavenumber at two axial locations, (a) $\Omega$ vs $\alpha_i$ at x*=5 mm and (b) $\Omega$ vs $\alpha_i$ at x*=30 mm. }
\label{fig:TripleSpatialAnalysis}
\end{figure}

\begin{figure}[h!]
\begin{center}
{\includegraphics[trim=2cm 1.5cm 2cm 2cm,clip,width=0.65\linewidth]{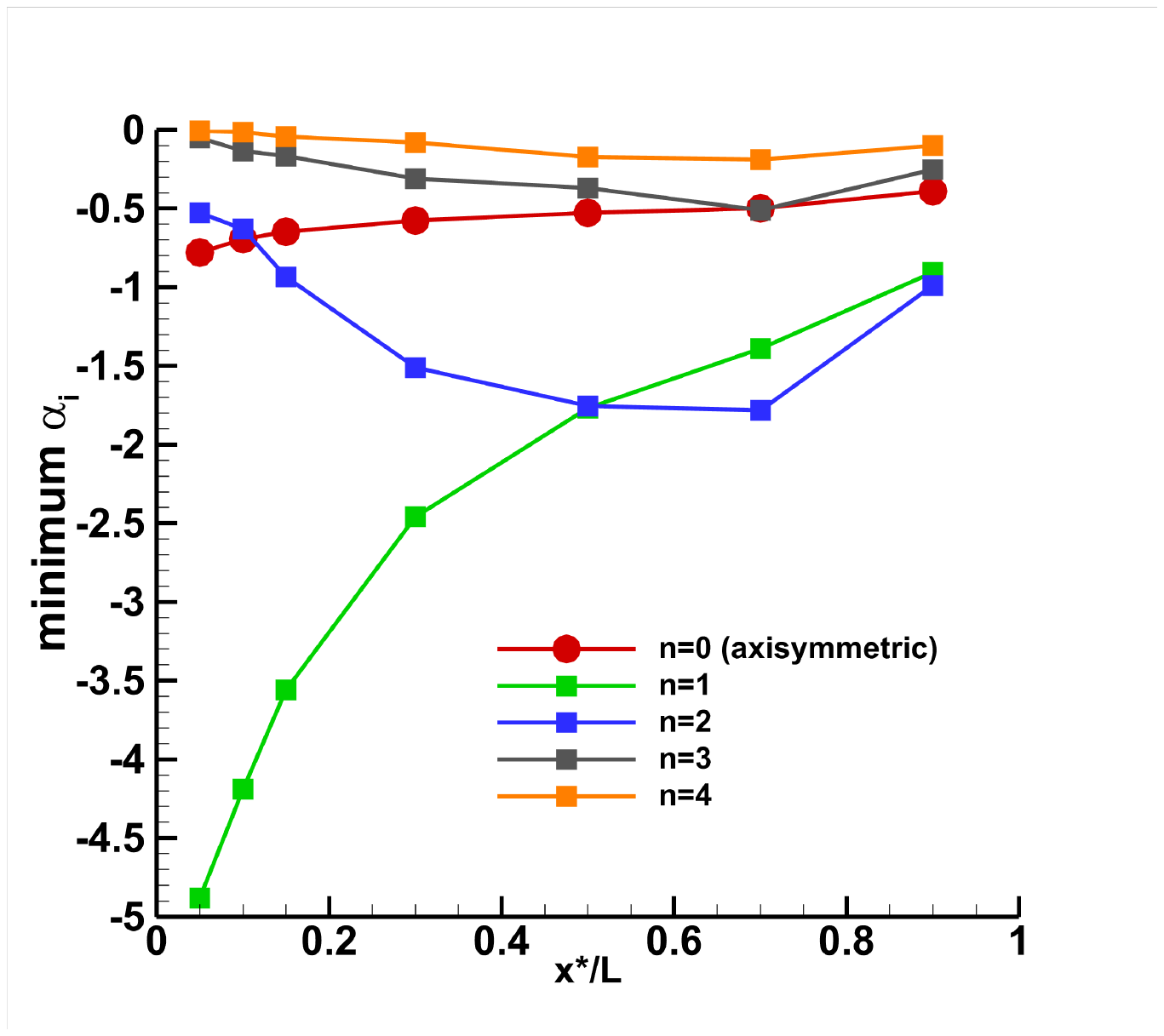}}
\end{center}
\caption{Most amplified azimuthal wavenumber along the cone surface.}
\label{fig:AzimuthalWavenumber}
\end{figure}

Using this triple deck approach, we analyzed the stability of the axisymmetric and non-axisymmetric disturbances at several locations on the cone. In the following results $a/r_s=0.87$, indicating the shock is very close to the surface of the cone. Solving first for spatial stability, Figure~\ref{fig:TripleSpatialAnalysis} shows the 
imaginary part of the solution, $\alpha_i$ of the most dangerous modes, {\em i.e.,} those that have the lowest minima value in $\alpha_i$ as established by Duck \& Hall~~\cite{DuckHall21990},  for each azimuthal wavenumber at two axial locations considering up to four axial modes, $n$.  
 It can be seen in Figs.~\ref{fig:omegaalfaix5mm} and \ref{fig:omegaalfaix30mm} 
  that the $n=1$ mode has the smallest minimum, meaning it is the most amplified or most dangerous mode.  It should be also noted that for all  frequencies $\Omega$  considered, there is always at least one non-axisymmetric mode that is amplified more than the axisymmetric mode ($n=0$),   This result is similar to the works of Cowley \& Hall~\cite{CowleyHall1990} and Seddougui \& Bassom~\cite{SeddouguiBassom1997} who found that for the 
case of an attached shock configuration 
 they obtained higher amplification rates for the non-axisymmetric modes 
 compared to a case where the shock was not included. In their work, the shock was included in the upper deck with an appropriate boundary condition, whereas the post shock values used as the boundary layer edge quantities and upper deck of the triple deck structure only consisted of the inviscid region for the cases that did not include the shock.   It can also be seen that the spatial growth rate of the $n=1$  mode decreases at the downstream location where additional modes ($n$ = 2, 3, 4 etc.)  start to show up and  the difference among the modal peak growth rates  becomes smaller.
  The curves also show similar characteristics to those given in Seddougui \& Bassom~~\cite{SeddouguiBassom1997}, where all  modes asymptote to zero as the frequency mode, $\Omega$, becomes large.  
 Using the largest amplification rate of each azimuthal wavenumber at every axial location, we obtain Fig.~\ref{fig:AzimuthalWavenumber} which shows that   the most dangerous azimuthal mode for the region close to the cone tip is $ n=1$.  Further downstream, beyond 50~mm (or $x^*/L$ = 0.5) from the cone tip, it can be seen that the  $n=2$ mode  has an amplification rate as high as the $n=1$ mode.  In contrast, the $n=3$ and higher modes are less amplified than the axisymmetric mode ($n=0$) for the length of the cone considered here.

\begin{figure}[h!]
\begin{center}
\subfigure[]{\label{fig:ndenx4mm}\includegraphics[trim=3.5cm 0.5cm 4.0cm 1.2cm,clip,width=0.40\linewidth]{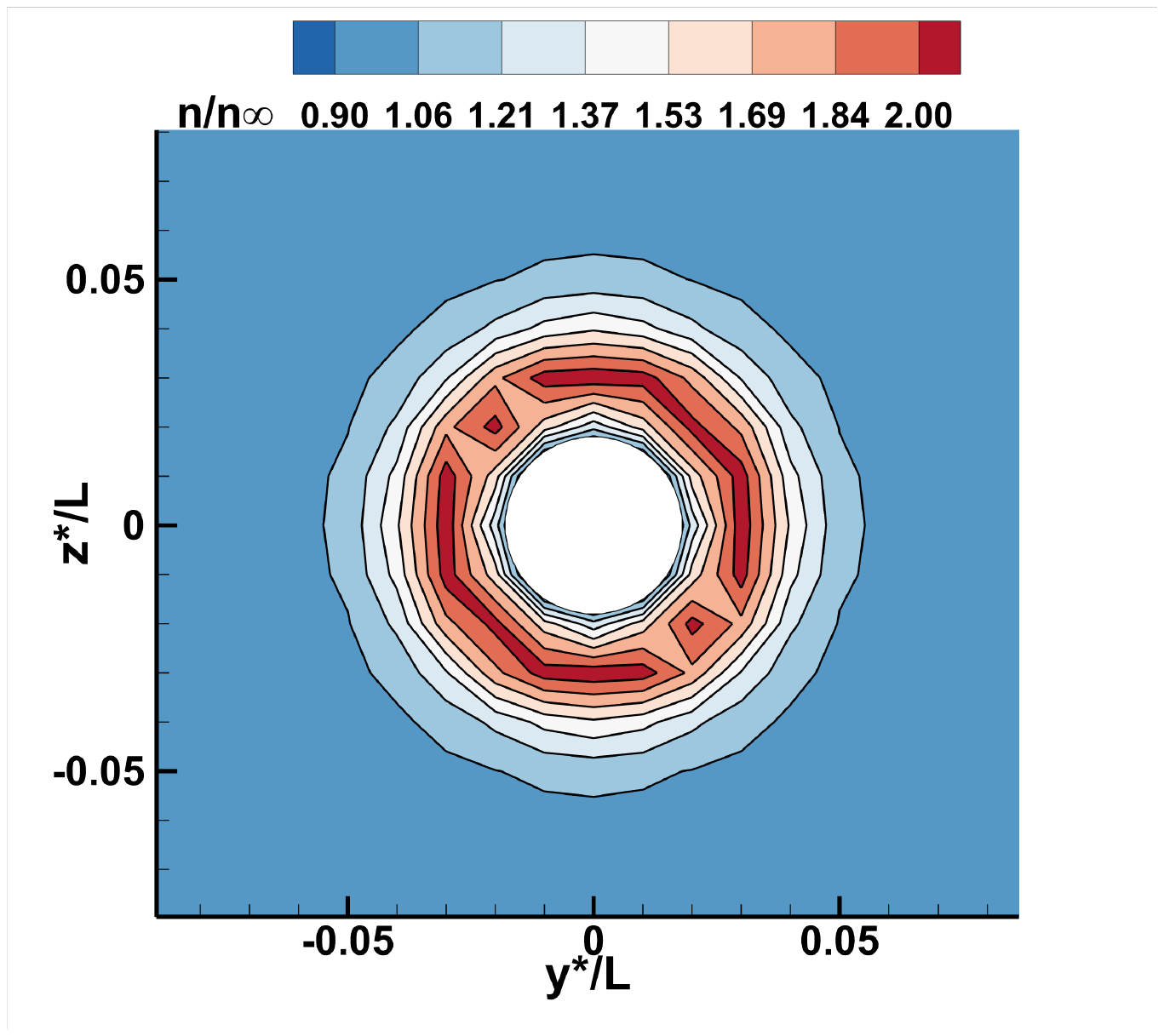}}
\subfigure[]{\label{fig:ndenx5mm}\includegraphics[trim=3.5cm 0.5cm 4.0cm 1.2cm,clip,width=0.40\linewidth]{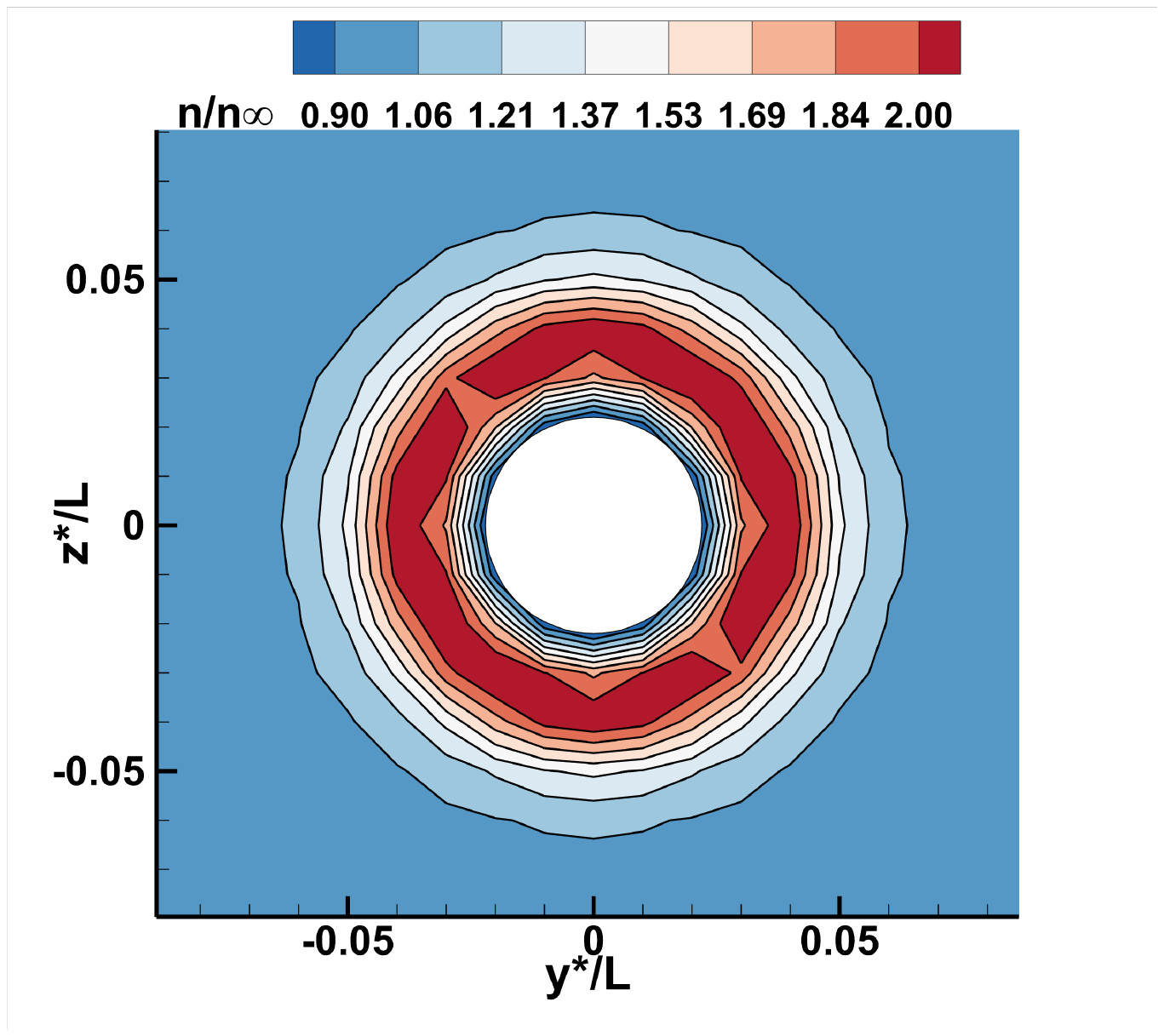}}
\subfigure[]{\label{fig:ndenx6mm}\includegraphics[trim=3.5cm 0.5cm 4.0cm 1.2cm,clip,width=0.40\linewidth]{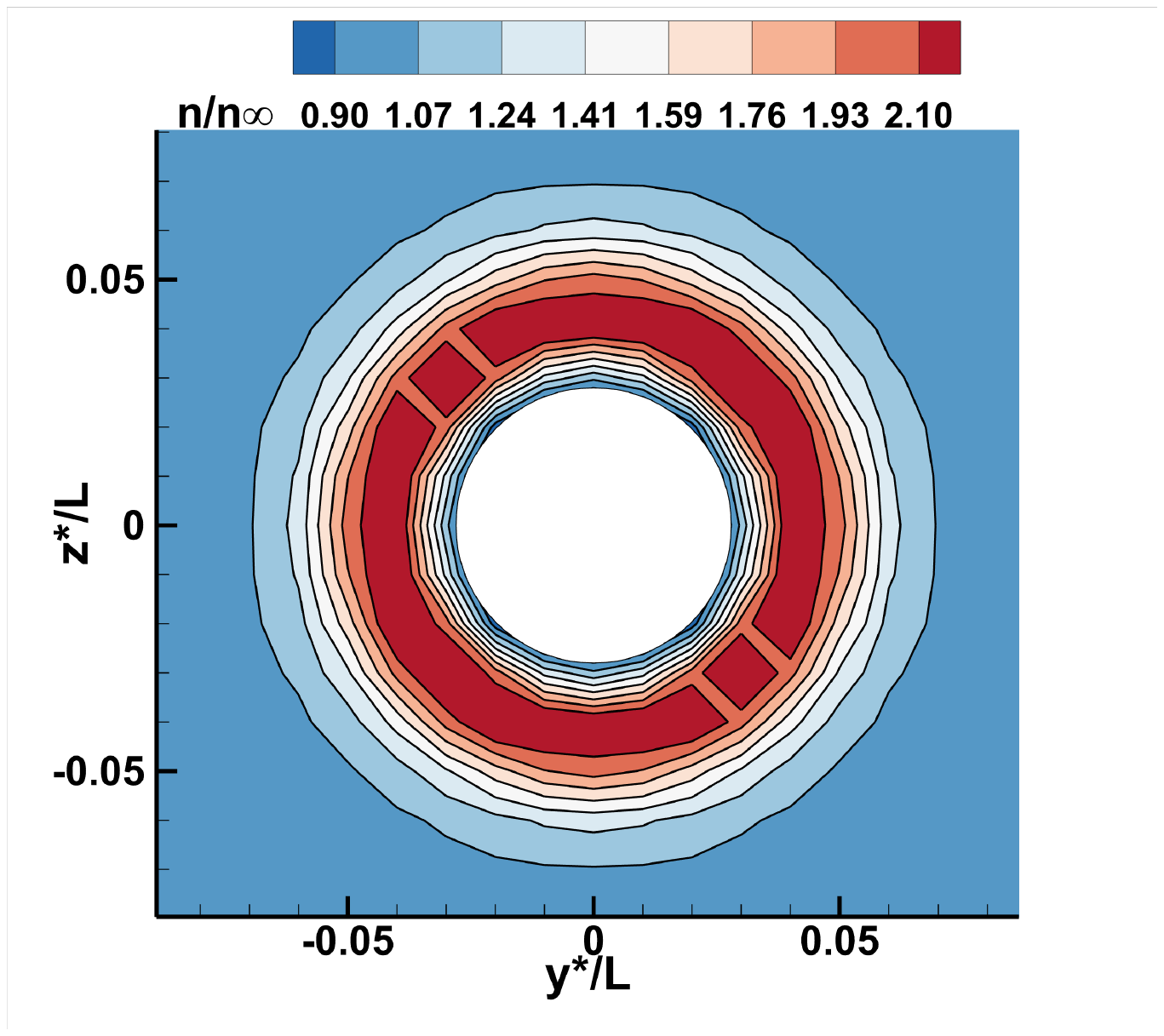}}
\subfigure[]{\label{fig:proloc}\includegraphics[trim=2.5cm 6.5cm 2.0cm 5.2cm,clip,width=0.40\linewidth]{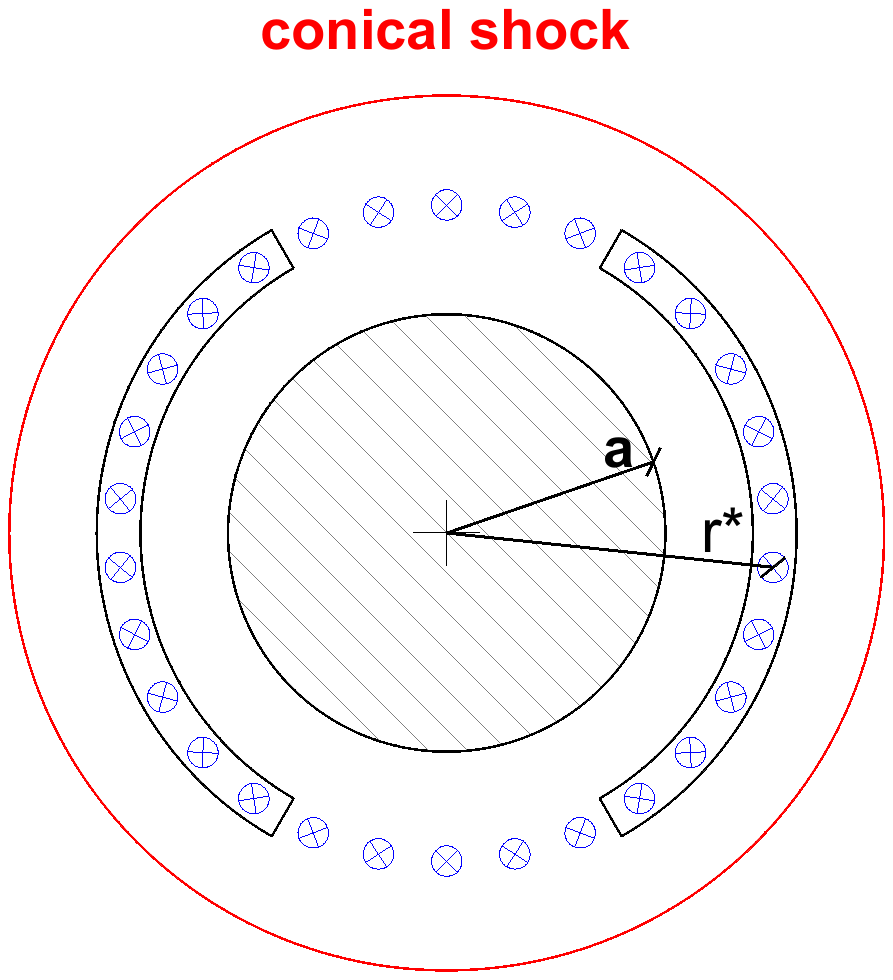}}
\end{center}
\caption{Number density contours looking head on at the tip showing $n=1$ at axial distances $x^*=4, 5,$ and 6~mm in (a), (b) and (c) respectively. Number density is normalized by the free stream number density.  Subfigure (d) shows the probe locations.}
\label{fig:NdenNonAxi}
\end{figure}

To verify this TDT/linear stability analysis, a fully three-dimensional DSMC simulation was performed over a single cone without the assumption of axial symmetry or the use of an azimuthal slice chosen based on an a-priori wavenumber~\citep{hao_fan_cao_wen_2022}. Figure~\ref{fig:NdenNonAxi} shows the instantaneous number density contours predicted by DSMC  where it can be seen that the flow is indeed non-axisymmetric with two large ``kidney-like'' structures present with a periodicity of $\phi=180^\circ$.  Based on  Eq.~\ref{eqn:perturbationansatz}, this is consistent with an azimuthal wavenumber disturbance of $n=1$, the major mode seen in Fig.~\ref{fig:AzimuthalWavenumber} which has an integer value as expected. The mode also seems to appear in the viscous region of the flow where the shock is very close to the cone surface  at the three selected axial locations, $x^*$ = 4, 5, and 6~mm, as shown in Figs.~\ref{fig:ndenx4mm},~\ref{fig:ndenx5mm} and~\ref{fig:ndenx6mm}.

\begin{figure}[h!]
\begin{center}
\subfigure[]{\label{fig:TDTempAmp}\includegraphics[trim=0cm 0cm 0cm 0cm,clip,width=0.48\linewidth]{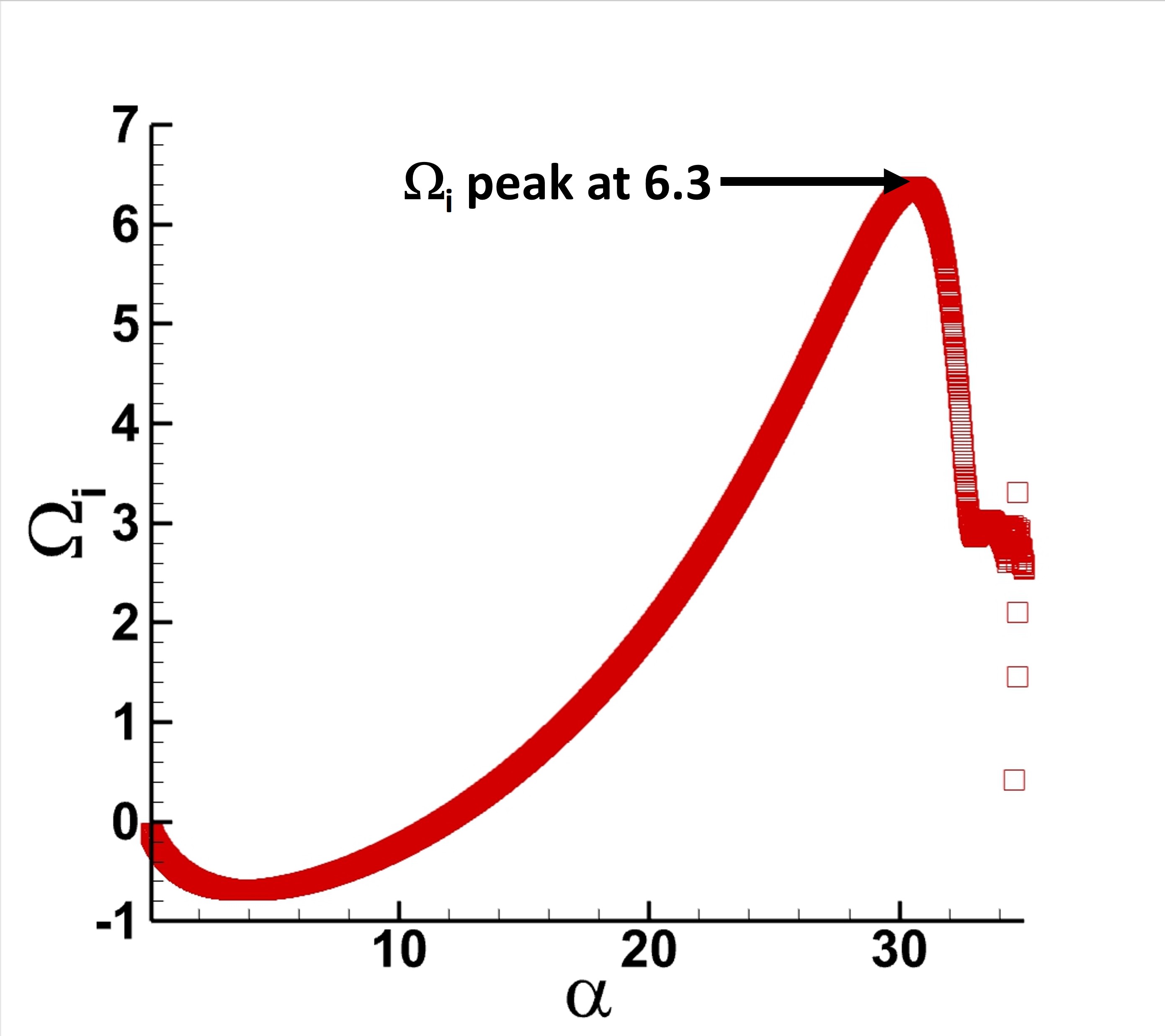}}
\subfigure[]{\label{fig:sigmaw}\includegraphics[trim=2.3cm 0.2cm 2.3cm 1.2cm,clip,width=0.48\linewidth]{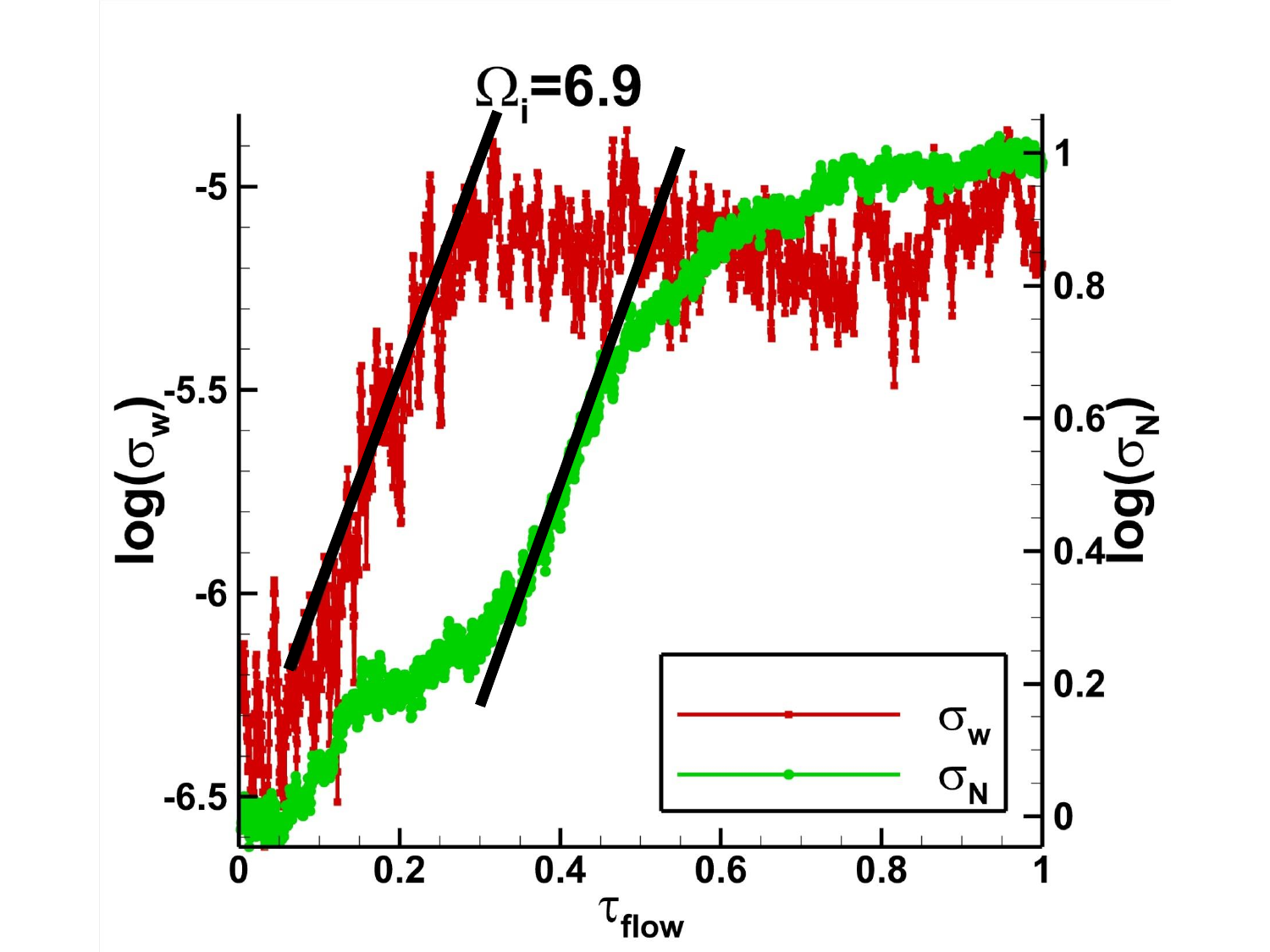}}
\subfigure[]{\label{fig:PSD3D}\includegraphics[trim=2.5cm 0.5cm 2.8cm 2.1cm,clip,width=0.48\linewidth]{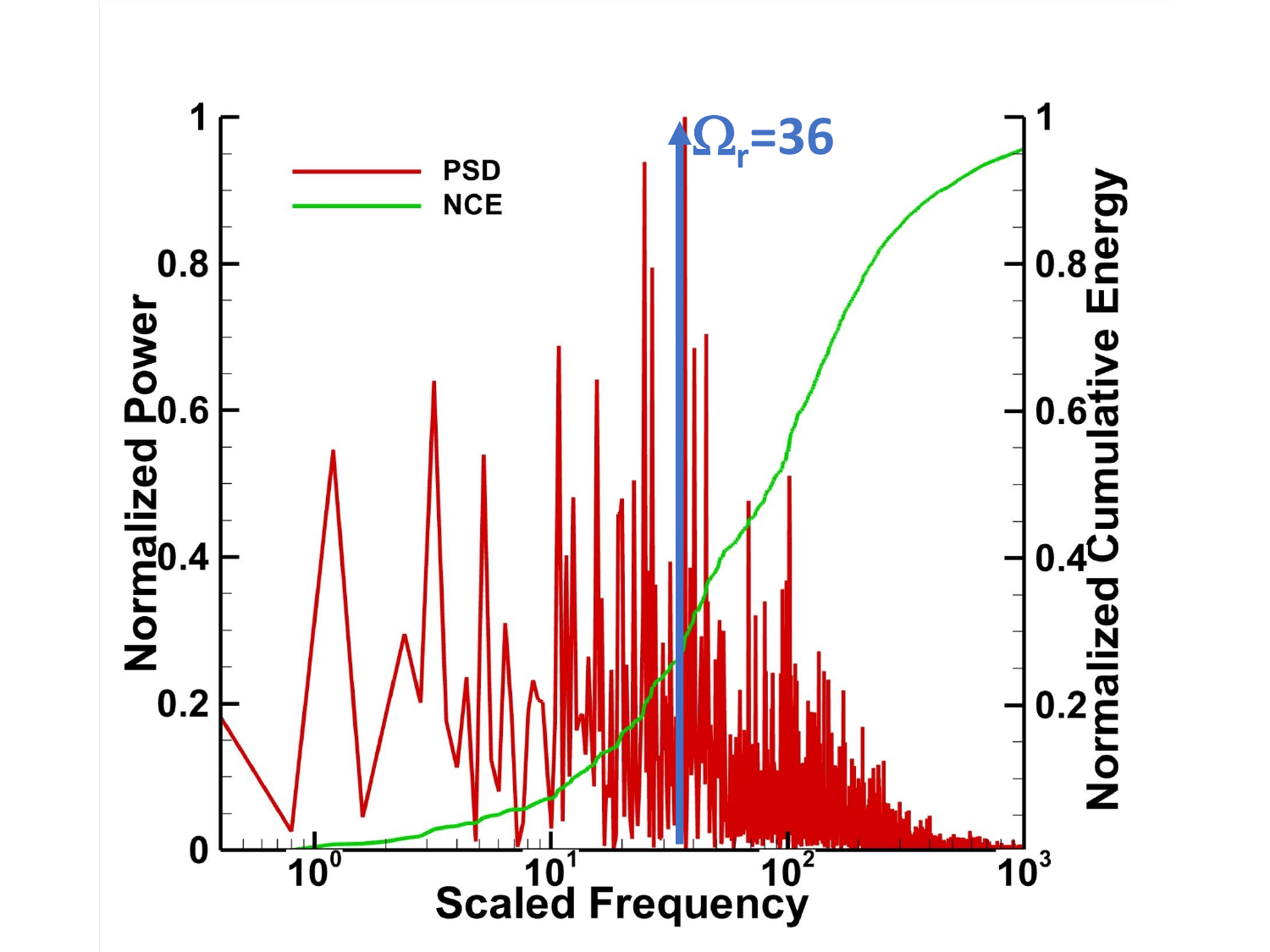}}
\end{center}
\caption{Temporal amplification and spectral content of the probe data at an axial location x$^*$=30 mm from the tip of the cone. $\Omega_r=36$ corresponding to the peak from the TDT. (a) Temporal amplification rate from TDT, (b) rms of the DSMC angular speed and DSMC number density deviation from the axisymmetric case, black lines show the fitted growth rate for these rms values and (c) PSD of the DSMC pressure signal.}
\label{fig:TripleDeckTemporal}
\end{figure}

A similar analysis for the temporal stability case (i.e. $\Omega$ complex, $\alpha$ real) at an axial distance from the cone tip of 30~mm for $n=1$ shows that there is a large range of wavenumbers that are amplified for $10 < \alpha < 30$, see Fig.~\ref{fig:TDTempAmp}.  As the figure shows, the highest amplification rate  was found to be $\Omega_i=6.3$ for the wavenumber ranges shown in the figure, thus suggesting that this  will also be the most probable rate to appear in the DSMC flowfield macroparameter data since it has the greatest range of wavenumbers that are amplified.  Using the time accurate numerical probe data from the DSMC simulation, the amplification rate from TDT analysis can be compared with that of the  root mean square of the azimuthal velocity $\sigma_w$ and the deviation of number density at a given location from the corresponding one in the  axisymmetric DSMC simulation, $\sigma_N$, as
\begin{eqnarray}
{\sigma _w} & = &\sqrt {\frac{1}{p}\sum\limits_1^p {{{\left( {\frac{w}{{{U_\infty }}}} \right)}^2}} }  \\
{\sigma _N} &=&\sqrt {\frac{1}{p}\sum\limits_1^p {{{\left( {\frac{{{N_{3D}} - {N_{axisymmetric}}}}{{{N_\infty }}}} \right)}^2}} }
\end{eqnarray} 
where $N$ is the number density, $p$ is the total number of probes and $w$ is the azimuthal velocity. A total of 32 probes were placed azimuthally  (see Fig.~\ref{fig:proloc})  within the flow layer where the broken structures formed.  As can be seen  in Fig.~\ref{fig:sigmaw} both $\sigma_w$ and $\sigma_N$ show amplification between normalized flow times of 0.2 to 0.4 with rates of $\Omega_i=6.9$, where the normalized flow time, $\tau_{flow}$ is,
\begin{equation}
{\rm{\tau_{flow}= }}\frac{{t^*{U_\infty }}}{L} \label{eqn:FlowTime}
\end{equation}

The figures also show that the triple deck theory predictions are in good agreement with the DSMC amplification rates.   For the temporal stability case, the dimensionless frequency corresponding to the wavenumber of the peak amplification rate was found to be $\Omega_r=33.2$ as shown in Fig.~\ref{fig:TDTempAmp}.  The power spectral density of the DSMC pressure data at this axial location shows that there is a low frequency broadband centered around $\Omega_r=36$, {\em i.e.,} the spectral content from the numerical probes matches the most amplified frequencies predicted by TDT  for $n=1$, as shown in Fig.~\ref{fig:PSD3D}.

Even though the effect of the leading edge shock on promoting three-dimensional instabilities was postulated~\citep{CowleyHall1990,SeddouguiBassom1997}, the breaking of axial symmetry close to the tip of a sharp cone has not been previously observed either through simulation or experiments, to the best of our knowledge.  The reasons why it is hard to capture these azimuthal modes numerically is because the simulations need to be truly three dimensional, and the internal structure of the shock layer needs to be resolved therefore requiring a kinetic approach. Both these factors result in high computational costs. In terms of experiments, observations are difficult because either flow visualizations normal to the free stream flow direction are required or several sensing elements needs to be placed azimuthally at the surface of the conical test article.  As already mentioned in Sec.~\ref{sec:intro},  the most relevant experiments in this regard were for converging shocks~\citep{Takayama1987}, and the preliminary experiments of Kennedy \textit{et al.}~\cite{McGilvrayDCScitech2023} for a double cone geometry.

\begin{figure}[h!]
\begin{center}
\subfigure[]{\label{fig:SCvsDCcut}\includegraphics[trim=2cm 1.5cm 4cm 2cm,clip,width=0.48\linewidth]{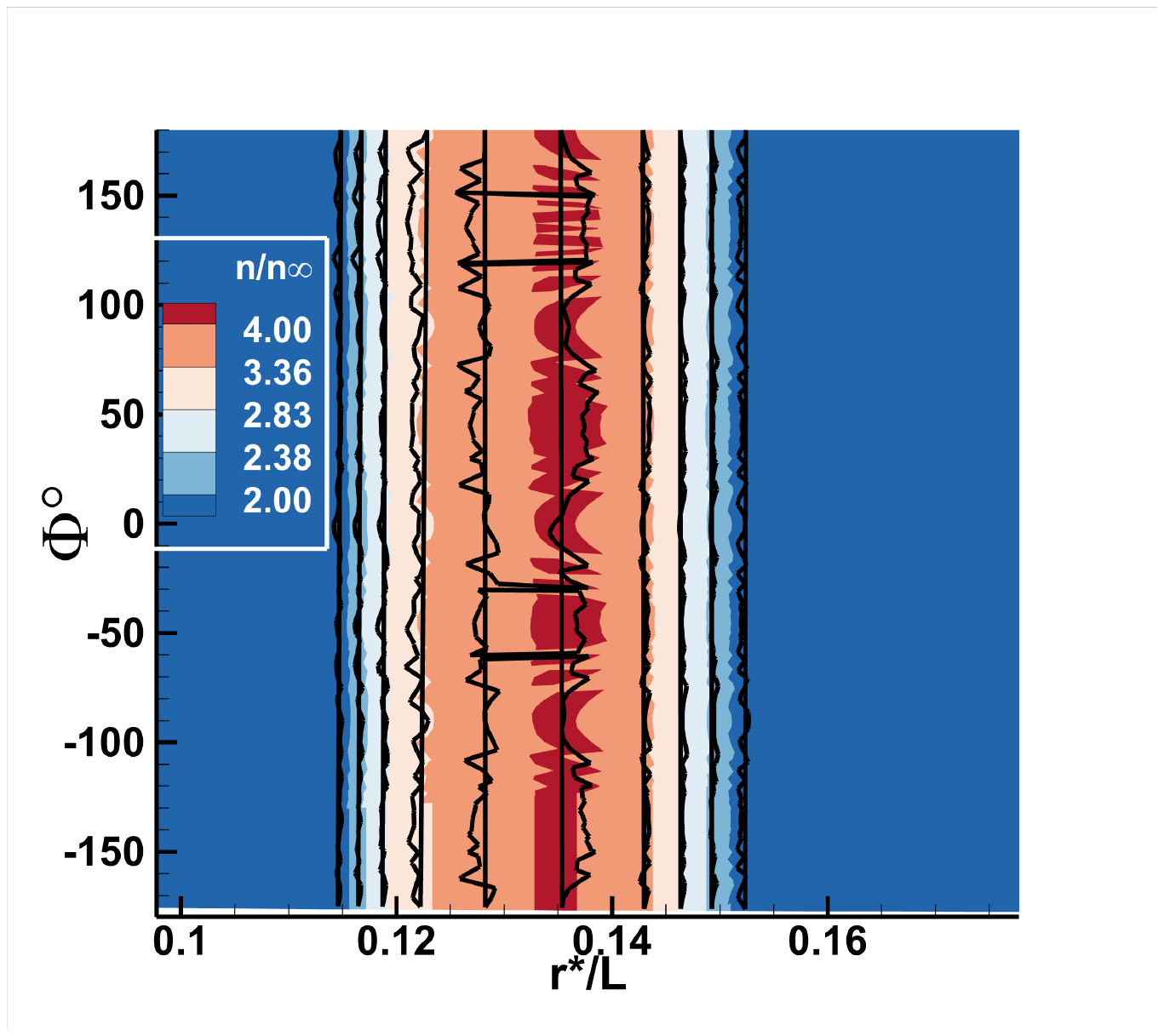}}
\subfigure[]{\label{fig:AzimuthalSpeed}\includegraphics[trim=2cm 1.25cm 4cm 2cm,clip,width=0.48\linewidth]{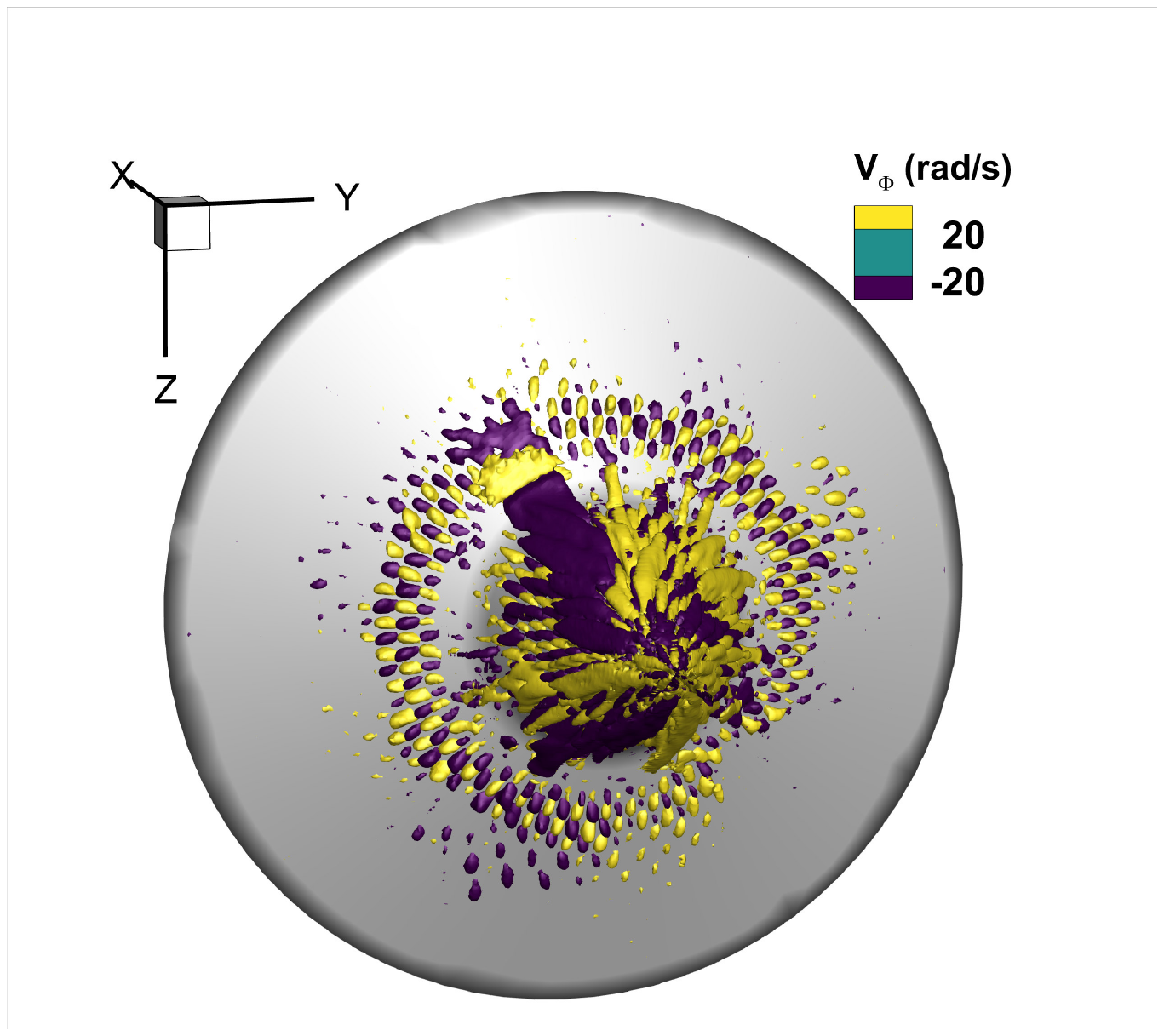}}
\end{center}
\caption{ (a) Normalized number density contours showing the $n=1$ mode for the single vs. double cone simulations (flooded vs. line contours, respectively) in the same streamwise plane 2.5~cm from the tip of the cone in cylindrical coordinates.  (b) Isosurfaces of angular velocity over a double cone.}
\label{fig:NumericalChecksforMode}
\end{figure}

Since the phenomenon of loss of axial symmetry close to the tip of a cone has not previously observed we explored additional flow and geometry conditions to test the robustness of the presence of the $n=1$ instability. For example, when the half angle of the base case was changed slightly, {\em i.e.} $1-2^{\circ}$, we could still observe the $n=1$ mode,  demonstrating the robustness of the mode. In fact, we confirmed that this mode  indeed originates from the tip of the first cone and is not affected by the geometry downstream by checking at the same axial locations for both the single and double cone simulations that it appears at the same azimuthal locations.  
 Converting the DSMC Cartesian macroparameters to cylindrical coordinates, {\em i.e.,}
\begin{eqnarray}
r = \sqrt {{y^2} + {z^2}}, \Phi  = \arctan (z/y) \\
{V_\varPhi } = \frac{{ - y{V_z} + z{V_y}}}{{{y^2} + {z^2}}}
\end{eqnarray}
where the superscript ``*'' indicating a dimensional quantity has been suppressed,  Fig.~\ref{fig:SCvsDCcut} shows the unstable mode appearing at the same azimuthal value.  

Another important question to consider is how 
the two kidney-like structures shown in Fig.~\ref{fig:NdenNonAxi}  change temporally and spatially as the flow evolves.  
For the time snapshots shown, the non-axisymmetric structures seem to be are arranged at the same azimuthal positions in the different streamwise planes given in Fig.~\ref{fig:NdenNonAxi}. However, much further downstream it was observed that these non-axisymmetric structures evolve and rotate in space.  The fact that there does not seem to be any sensible change in the disturbance locations for these different time snapshots is because 
 the calculated angular rotational speeds are at a maximum 20~rad/s, as shown in Fig.~\ref{fig:AzimuthalSpeed}, where the azimuthal velocities were also calculated using the above equations.  This non-zero angular rotation speed, also a proof of axial symmetry breakdown, corresponds to only a 
 $1-2^{\circ}$ of rotation for a $1~ms$ time period.   Since this is the total physical run time for a DSMC computation, selected to be similar to those of  experiments at these conditions, it is not expected that there would be any significant rotation in these structures in that time period.  
 
Finally, we  investigated whether axial symmetry would be lost for similar, but different, free stream conditions for flow over a single cone.  We performed DSMC simulations at two other conditions, {\em i.e.,} M$_\infty=6$ and Re$_\infty=1\times10^{4}~m^{-1}$ and M$_\infty=16$ and Re$_\infty=1\times10^{4}~m^{-1}$.  For the first case, 
we did not observed any three dimensionality in the single cone simulations. When compared with the appropriate scalings given in the S\&B, it was found that this case did not satisfy the limiting parameters that require the boundary layer thickness to be on the same order as the distance of the shock from the surface (the former was too thick).  Also, since the Mach number was smaller compared to our baseline case, the shock was further away from the cone surface and did not amplify the non-axisymmetric modes. The second case  did not show any three dimensionality either, however the reason was different.  In this second case, the shock is quite diffuse which reduced the related gradients to such a level that there was no amplification of any  non-axisymmetric modes.

\pagebreak
\section{Three dimensionality in the double cone system}\label{sec:DoubleConeStability}

In contrast to a single cone, the double cone creates several high gradient layers starting with the conical shock (CS) due to the sharp leading edge. As Fig.~\ref{fig:SchematicDC} summarizes, flow develops over the first cone and separates because of the effect of the shock created by the second cone creating a shock induced separation region. Due to this separation region several complex flow structures appear such as the separation shock (SS), bow shock (BS), separated shear layer (SL), and lambda shock-lets. At the corner, expansion waves (EW) are also present as the flow expands again. Note that the TDT analysis used in Section~\ref{sec:singleConeStability} cannot be applied to the double cone configuration because  it does not account for flow separation or additional shocks.  

\begin{figure}[h!]
\center
\subfigure[]{\label{fig:breakdown}\includegraphics[trim=2.5cm 0.5cm 2.8cm 2.1cm,clip,width=0.48\linewidth]{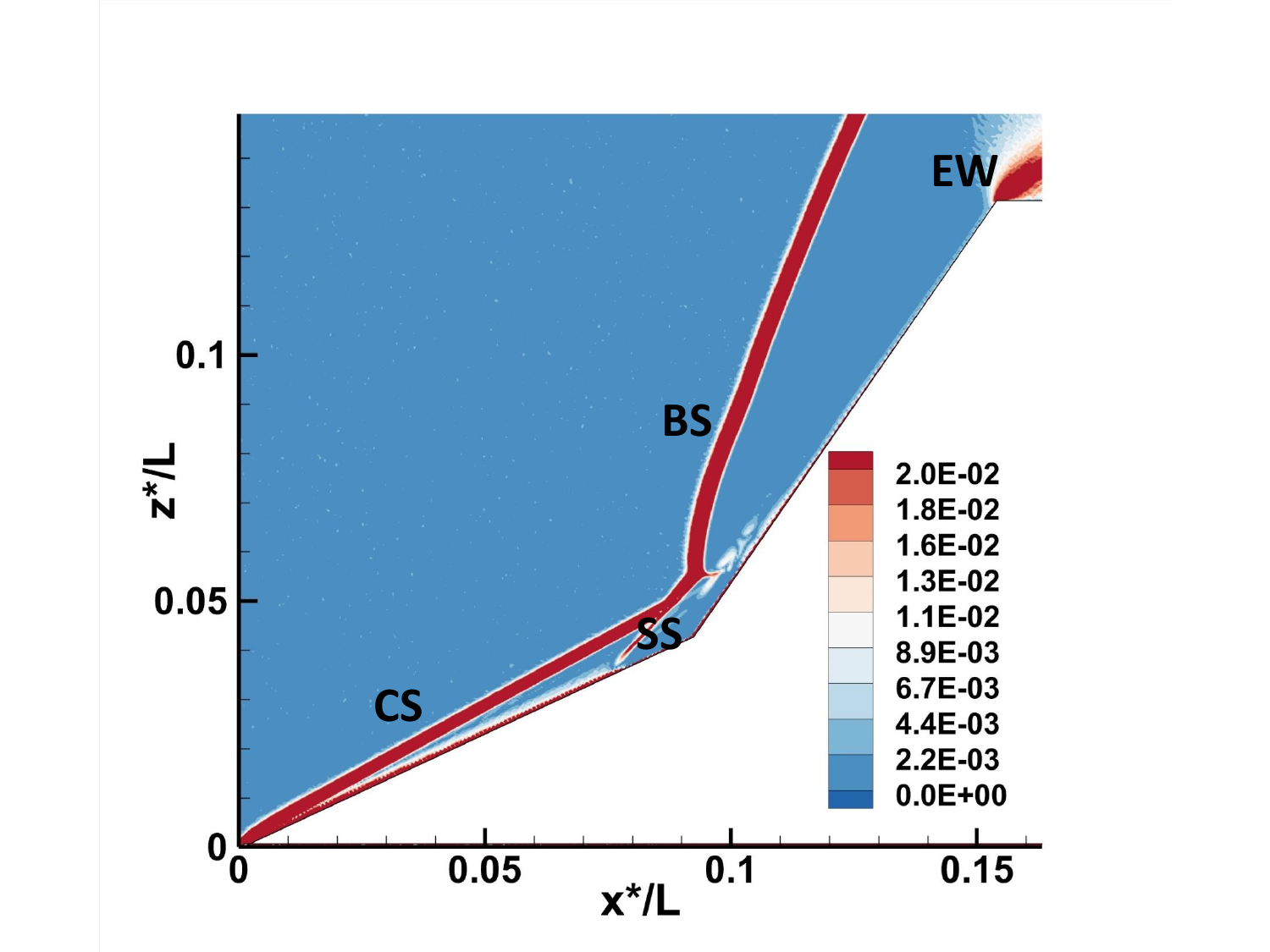}}
\subfigure[]{\label{fig:FlowRevAxi3D}\includegraphics[trim=2.5cm 0.5cm 2.8cm 2.1cm,clip,width=0.48\linewidth]{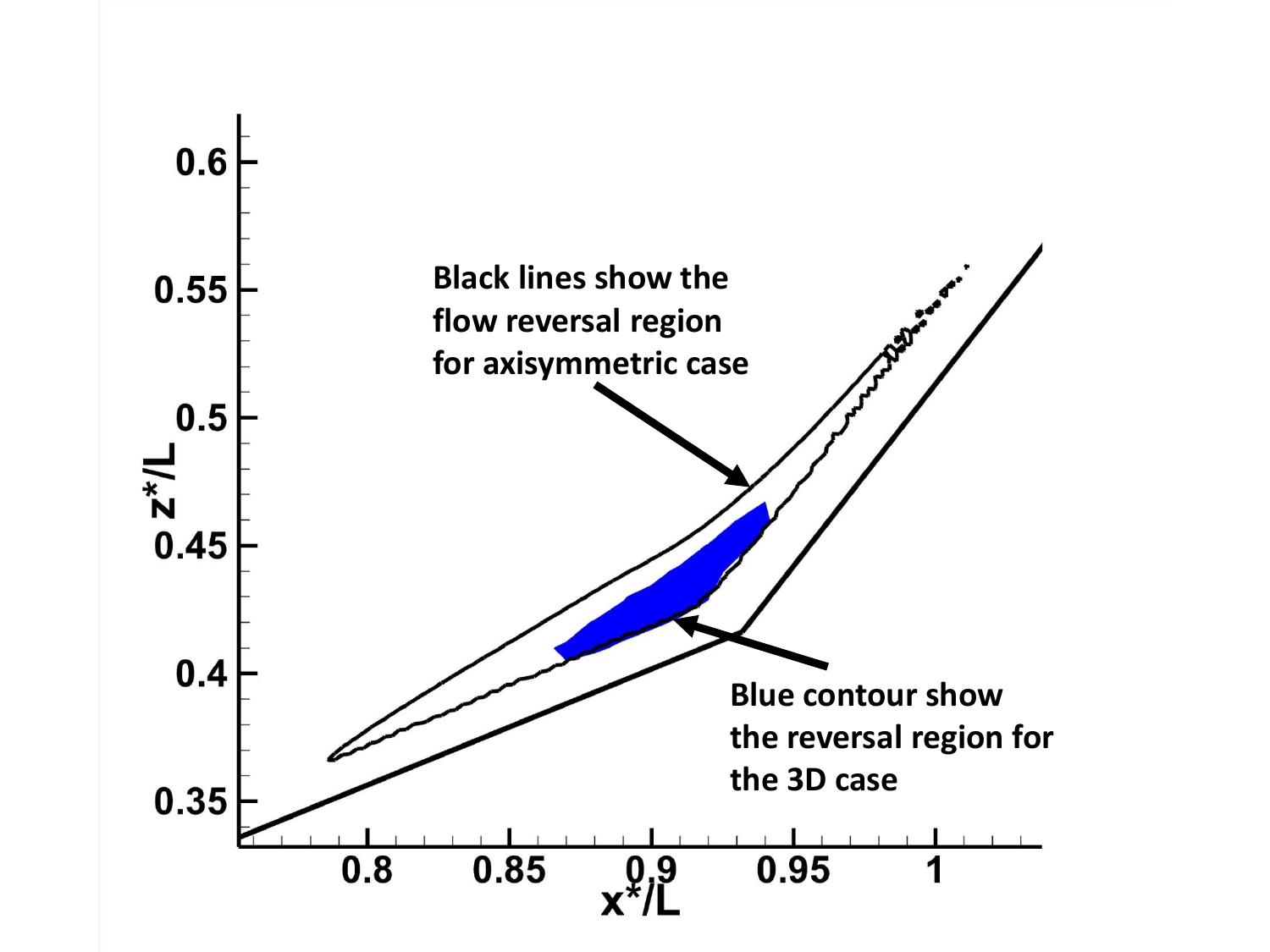}}
\caption{(a) continuum breakdown parameter contours and (b) flow reversal comparison of 3D vs axisymmetric cases.}
 \label{fig:Axivs3DSch}
\end{figure}

As mentioned earlier, an important consideration of flows dominated by shocks, expansions and shear layers is the phenomena of continuum breakdown, or regions in the flow where the continuum assumption is not valid {\it regardless} of the free stream number density or the scale of the geometry. Bird~\cite{Bird} defined a continuum breakdown  parameter ($P$) that depends on density gradient, Mach number, and local mean free path given as;
\begin{equation}
\label{eqn:breakdown}
P = \sqrt {\frac{{\pi \gamma }}{8}} \frac{{M\lambda }}{\rho }\left| {\frac{{d\rho }}{{d\overline x }}} \right|
\end{equation}
and  suggested that if the local $P>0.02$, the continuum assumption is no longer valid.  This implies that continuum breakdown can occur not only for rarefied cases but flows with high gradient regions such considered in this work. Figure~\ref{fig:breakdown} shows that $P>0.02$ (dark red colored regions) in the conical shock, bow shock, separation shock and the expansion waves. This result indeed argues for the use of a kinetic method such as DSMC to capture the complex physics of the flow, as was demonstrated in the earlier work of Tumuklu \textit{et al.}~\cite{tumukluPoF,tumukluPoF2} for the same conditions. 

The characteristics of the separation region are also an important focus of the unsteady behavior of the double cone shock system.  As Fig.~\ref{fig:FlowRevAxi3D} shows, the size of the separation region that we predict is significantly different for an axisymmetric versus a full three-dimensional simulation, as was also observed by Hao \textit{et al.}~\cite{hao_fan_cao_wen_2022} for the same geometry but different free stream conditions.   
The separation region is much smaller for the 3D case compared to the axisymmetric case (see Fig.~\ref{fig:FlowRevAxi3D}). It can be seen that the separation bubble is not a closed recirculation region for the 3D case, clearly indicating that there are three dimensional effects present in that region.

\begin{figure}[h!]
\begin{center}
\subfigure[]{\label{fig:ch3D}\includegraphics[trim=0cm 0cm 0cm 0cm,clip,width=0.48\linewidth]{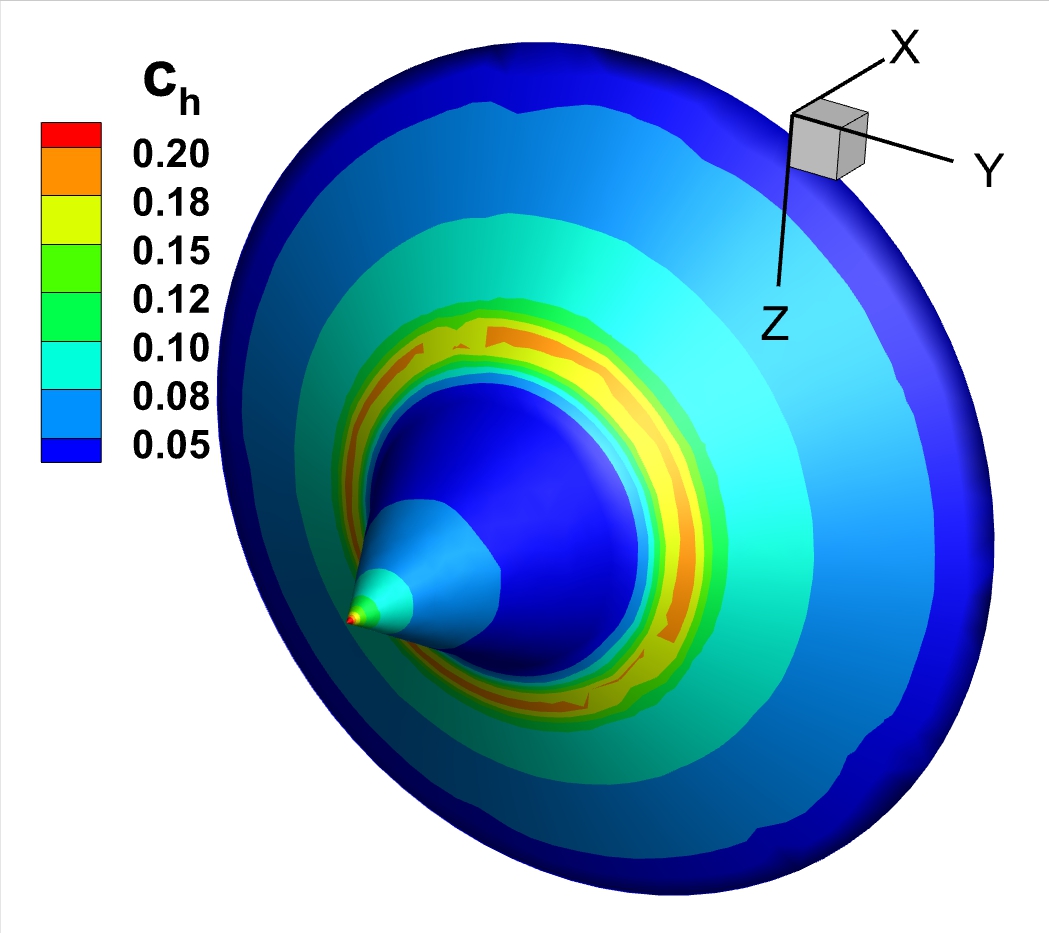}}
\subfigure[]{\label{fig:cf3D}\includegraphics[trim=0cm 0cm 0cm 0cm,clip,width=0.48\linewidth]{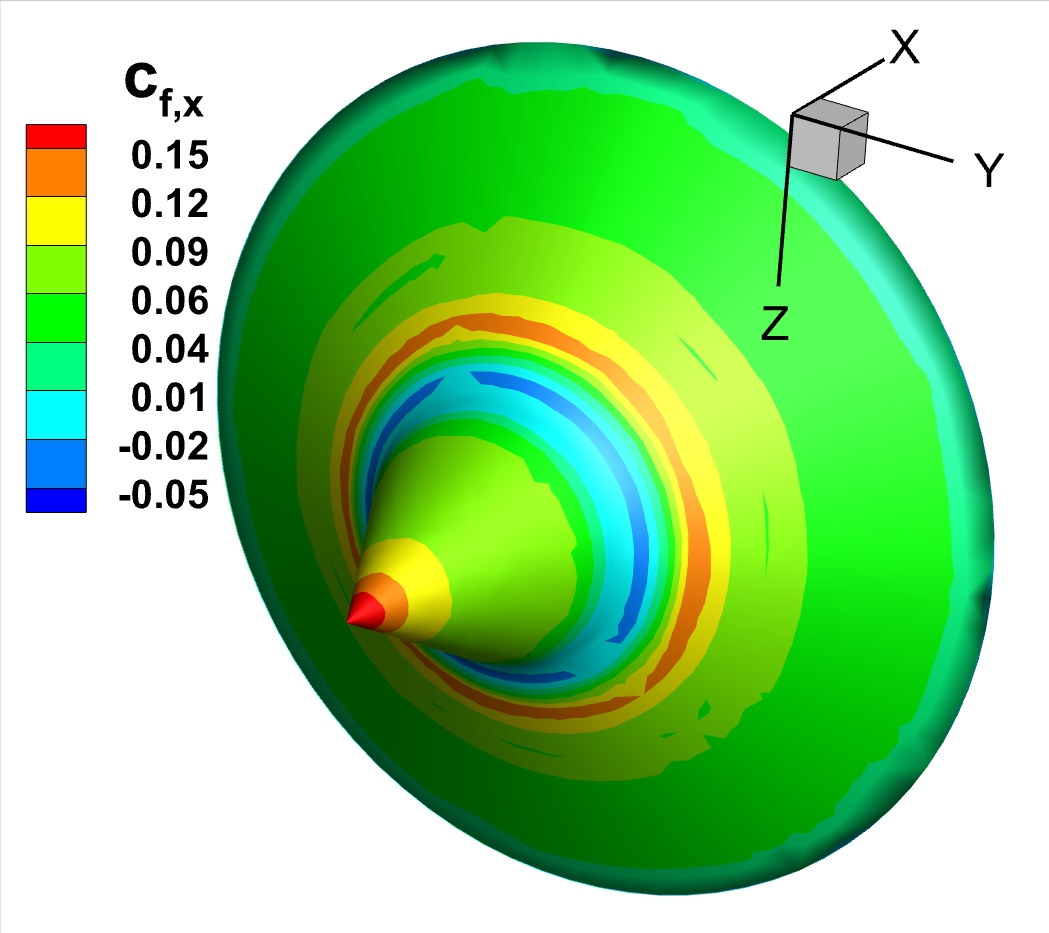}}
\subfigure[]{\label{fig:cfch3D}\includegraphics[trim=2cm 1cm 2cm 2cm,clip,width=0.48\linewidth]{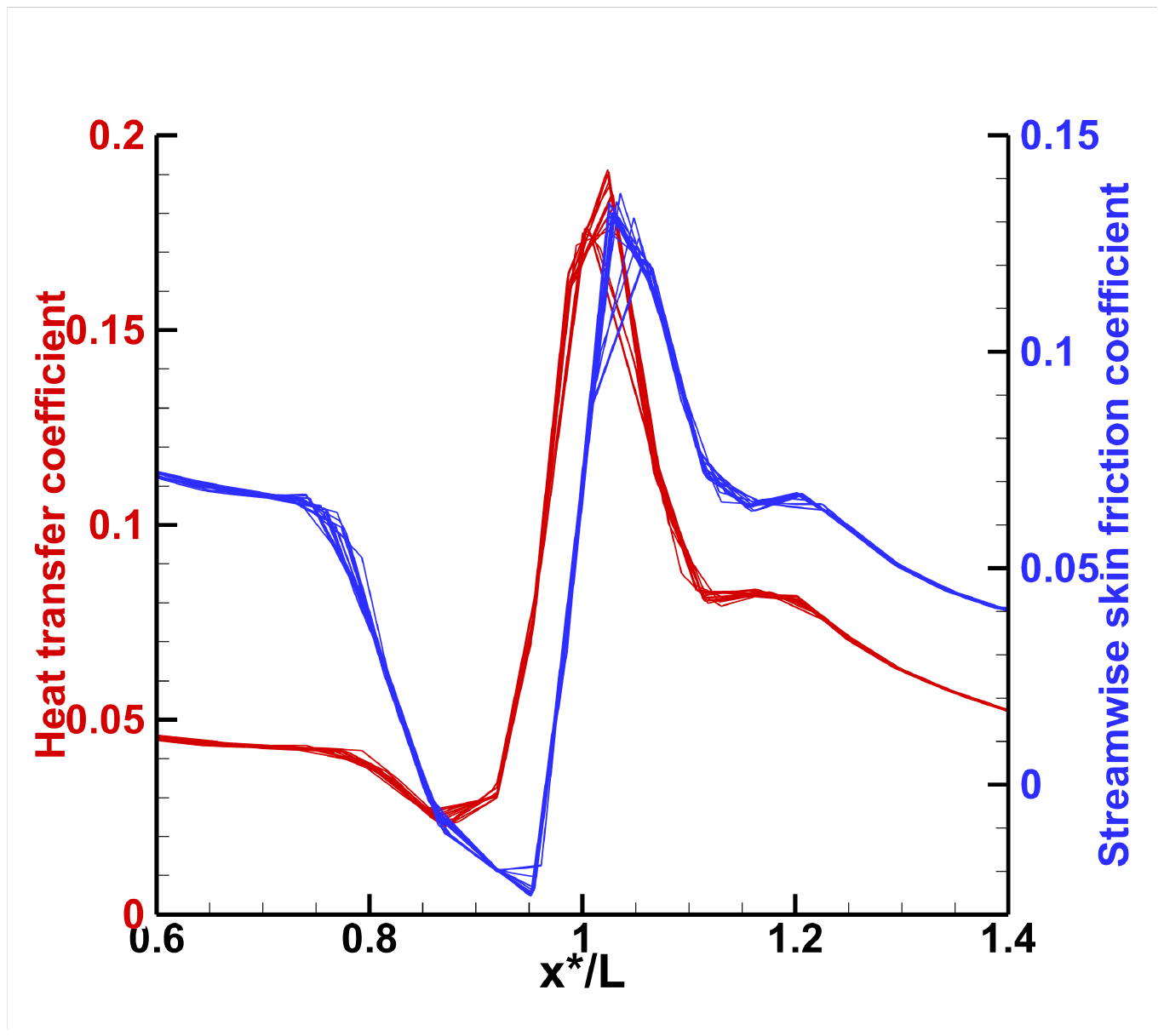}}
\subfigure[]{\label{fig:cf3Dcloseup}\includegraphics[trim=0cm 0cm 0cm 0cm,clip,width=0.48\linewidth]{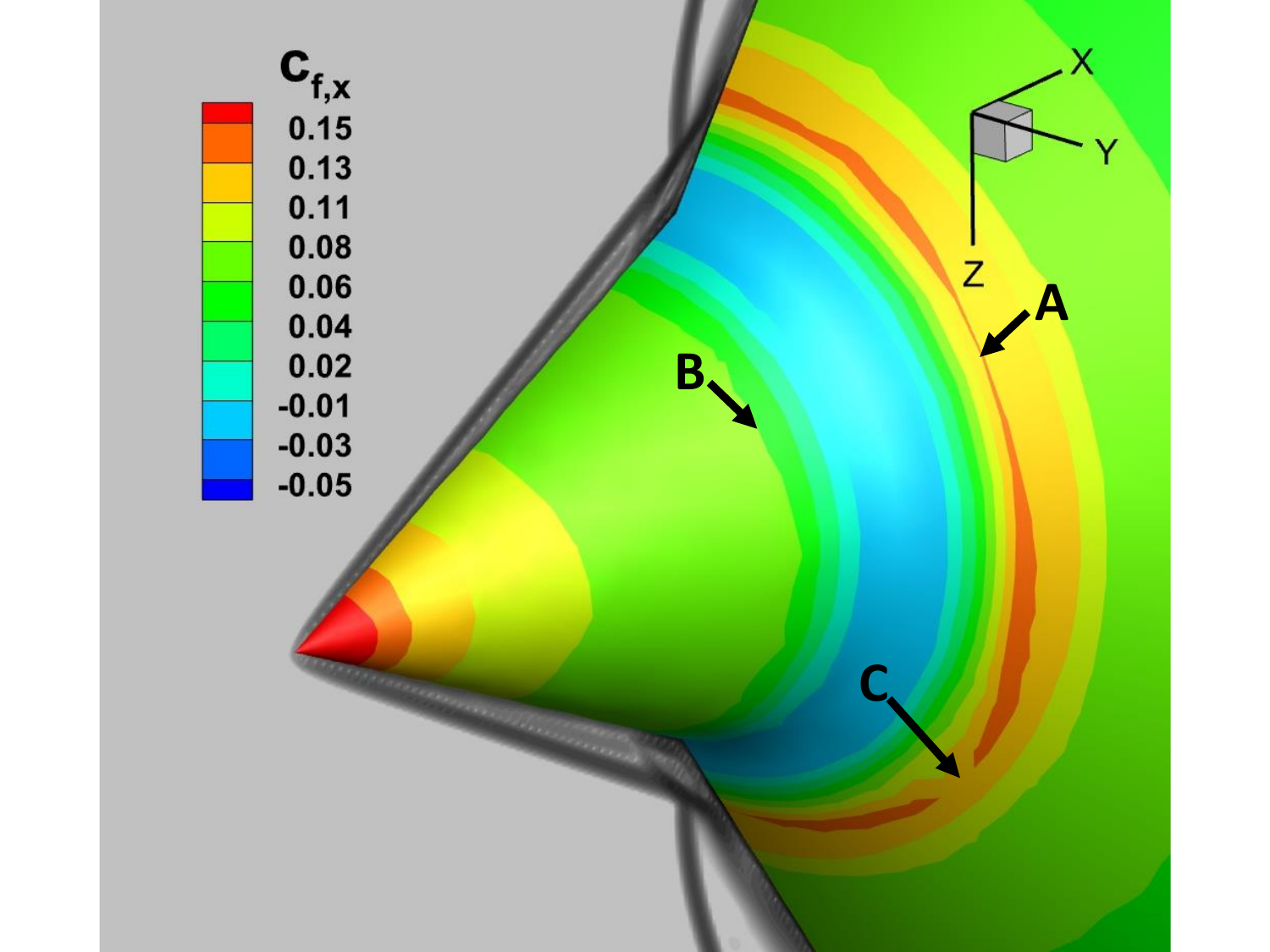}}
\end{center}
\caption{Surface properties of a three-dimensional double cone with  detailed views in the separation region. (a) Heat transfer coefficient (c$_h$), (b) streamwise skin friction coefficient (c$_{f,x}$), (c) heat transfer and streamwise skin friction coefficient along the axial direction. 10 different lines along the surface of both cones showed. Note that these lines overlap in the undisturbed regions but several lines can be seen near separation reattachment, (d)  zoomed view at the cone junction of a simulated Schlieren  with the streamwise skin friction coefficient.  Note that locations A, B, and C are  the location of the transmitted conical shock, wavy separation line,  and the discontinuity in the circular shape for the 0.12 contour level occur, respectively.}
\label{fig:SurfaceProperties3D}
\end{figure}

To further investigate three dimensionality of the flow in the separation bubble, we consider the predicted surface parameters from a full three-dimensional simulation in the conical shock  region to determine if there is any indication of non-axial symmetry.  It can be seen from Figs.~\ref{fig:ch3D} and \ref{fig:cf3D} that axial symmetry is broken especially near the separation and reattachment locations for both heat transfer and streamwise skin friction coefficients with the difference of axisymmetric versus three-dimensional on the order of 20\% for instantaneous time snap shots.   This is also evident from Fig.~\ref{fig:cfch3D} which shows the streamwise skin friction and heat transfer coefficients for several lines along the surfaces at different azimuthal angles, {\em i.e.,} for a perfectly axisymmetric case all curves should overlap. More interestingly, we observed that there are two large structures appearing in both the heat transfer and skin friction contours at the location where the transmitted conical shock impinges on the wall.  The shock structure in a planar cut normal to the azimuthal direction shown in Fig.~\ref{fig:cf3Dcloseup} demonstrates that this additional structure is due 
 to the loss of axial symmetry in the conical shock close to the cone tip itself, as discussed in the previous section,  that propagates downstream to the separation region.

\section{Conclusions}\label{sec:conc}

Hypersonic flows over conical shapes were modeled using DSMC to study the contribution of  three dimensionality to unsteady flow disturbances.  Starting with a simple cone, the triple deck formulations in combination with the linear stability of S\&B was used to predict 
azimuthal eigenmodes of this flow.   For our conditions, the theory showed that strongest amplification rate occurred for an azimuthal wavenumber of $n=1$ in regions quite close to the tip of the cone  due to the proximity of the conical shock to the viscous shear layer where non-axisymmetric modes are amplified through linear mechanisms.  Further comparison of the triple deck linear stability calculations showed that in addition to the azimuthal wavenumber, both temporal content and the amplification rate of these non-axisymmetric disturbances agree  well with the time accurate DSMC flowfield data.  To our knowledge, this is the first verification of triple deck theory by an exact kinetic method important because the latter has no a priori assumptions.

In addition to the loss of axial symmetry observed at the conical shock, the effect of axial symmetry assumptions on  the more complex shock-shock and shock-boundary layer interactions of a flow over a double cone were considered.   The results for the separation region showed that axisymmetric and three dimensional simulations differ in almost all of the main  flow structures. Three dimensional flowfields resulted in a smaller separation bubble with weaker shocks. Three dimensional effects  were manifest in the variation in surface parameters in the azimuthal direction as well.   Interestingly, the DSMC simulations showed that  the loss of axial symmetry in the separation region, begins at the cone tip.

\section*{Acknowledgments}
The research conducted in this paper is supported by the Office of Naval Research under Grant No. N000141202195 titled “Multi-scale modeling of unsteady shock-boundary layer hypersonic ﬂow instabilities,” with Dr. Eric Marineau as the Program Officer. This research is also supported by NSF ACCESS (previously XSEDE) Frontera supercomputer with the project number CTS22009. The authors are grateful to Dr. Vassilis Theofilis  for fruitful discussions.

\pagebreak
\bibliography{DCSC_main}



\end{document}